# Transient analysis of arm locking controller

**Yi Zhang,[1] Mingzhe Li,[1] Tong Wang,[1] Xinyi Zhao,[2] Long Ma,[2*] Shaobo Fang,[3,4*] and Ming Xin[1,5*]**

[1] *School of Electrical and Information Engineering, Tianjin University, Tianjin, 300072, China*
[2] *Sino-European Institute of Aviation Engineering, Civil Aviation University of China, Tianjin 300300, China*
[3] *University of Chinese Academy of Science, Beijing, 100049, China*
[4] *Institute of Physics, Chinese Academy of Sciences, Beijing, 100190, China*
[5] *Tianjin Key Laboratory of Brain-Inspired Intelligence Technology, Tianjin, 300072, China*
**email: longma@cauc.edu.cn, shaobo.fang@iphy.ac.cn, xinm@tju.edu.cn*

**Abstract:** Arm locking is one of the key technologies to suppress the laser phase noise in spaced-based gravitational waves observatories. Since arm locking was proposed, phase margin criterion was always used as the fundamental design strategy for the controller development. In this paper, we find that this empirical method from engineering actually cannot guarantee the arm locking stability. Therefore, most of the advanced arm locking controllers reported so far may have instable problems. After comprehensive analysis of the single arm locking's transient responses, strict analytical stability criterions are summarized for the first time. These criterions are then generalized to dual arm locking, modified-dual arm locking and common arm locking, and special considerations for the design of arm locking controllers in different architectures are also discussed. It is found that PI controllers can easily meet our stability criterions in most of the arm locking systems. Using a simple high gain PI controller, it is possible to suppress the laser phase noise by 5 orders of magnitude within the science band. Our stability criterions can also be used in other feedback systems, where several modules with different delays are connected in parallel.

## 1. Introduction

The observation of gravitational waves (GWs) [1] has opened a new window for humans to explore the unknown universe. To overcome the limitations from seismic gravity-gradient noise [2] in ground-based GW observatories, several spaced-based GW observatory projects: LISA [3], DECIGO [4], Taiji [5,6] and TianQin [6,7], have been initiated world widely in the past two decades, aiming to detect GWs in the frequency range from 0.1 mHz to 1 Hz (science band).

Similar to the ground-based GW observatory, spaced-based GW observatory is essentially a laser interferometer. For example, Taiji constellation consists of three spacecrafts oriented in an approximate equilateral triangle with 3 Gm arm length. Each spacecraft is equipped with two continuous wave (CW) lasers and coherent laser beams can be exchanged through six inter-spacecraft links. Heterodyne interferometry is employed at each spacecraft to extract the phase difference between laser signals that travel different link lengths. GWs will cause the three spacecraft to shift slightly with respect to each other, which can be measured from the phase error signals after heterodyne detection.

The scientific goal of these spaced-based GW observatories is to achieve a strain sensitivity of $10^{-21}/\sqrt{\text{Hz}}$ in the science band, which requires the frequency noise of the CW lasers better than $10^{-6}$ Hz/$\sqrt{\text{Hz}}$. However, even for the most stable free-running CW lasers, there is still 8 to 10 orders of magnitude gap to this extremely low noise level. In order to meet the strict requirement above, three techniques have been adopted [8]. The laser's frequency is first pre-stabilized to a fixed-length ultra-stable optical cavity using Pound-Drever-Hall (PDH) locking method [9]; then the arm length of the constellation, which is much stable than the laser's frequency in the science band, is used as a reference to further reduce the laser's phase noise, and this is called arm locking technique[10-27]; finally, the residual laser frequency noise can be canceled by time delay interferometry (TDI) [28-32], with the help of virtual delays introduced in data post processing.

As one of the crucial procedures in laser frequency noise suppression, the performance of arm locking directly determines the final detection sensitivity of the GWs. Therefore, since this technique was firstly proposed in [10], many efforts have been put into this field, both theoretically [11-22] and experimentally [23-27]. At the beginning, single arm locking was first investigated due to its simple structure [10-12]. Then dual arm locking was proposed to put the first null of the controller out of the science band [13], leading to a significant noise reduction in the high frequency range of the science band (e.g., [0.1 Hz, 1 Hz]). Furthermore, modified dual arm locking is presented [14] to solve the doppler-induced frequency pulling problem, while maintaining the gain advantages of dual arm locking. Recently, optical frequency comb was also introduced into arm locking [22], using optical frequency division, all the intrinsic nulls of the single arm locking sensor can be eliminated, resulting a good noise suppression performance within the entire science band.

Although remarkable progress has been achieved for arm locking in the past two decades, all the arm locking controllers reported so far were optimized in the frequency domain, based on the well famous phase margin criterion in engineering. However, there is no rigorous mathematical derivation to prove that this criterion is always valid. Besides, because of the arm length delay, the transient peaks arise during the locking start time can periodically reinject into the system, which cannot be modeled by the frequency domain analysis, as the latter can only provide the system's steady state response. [11] simply assumed that these periodical transient peaks will decay to zero after sufficiently long time, yet in fact this is not always true if the controller was not carefully designed. To solve the problems above, in this paper, a comprehensive transient analysis of the arm locking controller will be given for the first time. The evolution of those transient peaks will be analytically derived. Based on these results, an exact mathematical criterion for the stability of a general arm locking controller will be presented.

The remaining of this paper will be organized as below: In Section 2, the transient response of single arm locking will be first discussed, then a stability criterion for a general single arm locking controller will be given. This criterion will be generalized to dual arm locking, modified-dual arm locking and common arm locking controllers in Section 3, 4 and 5, respectively. Two controllers in literature was also checked by our criterions in Section 6, before we conclude in Section 7.

## 2. Single arm locking

As single arm locking has the simplest feedback structure, we will first analyze its transient response. Fig. 1(a) gives the diagram of single arm locking control system, where $\varphi_0(t)$ is the original laser phase noise in time domain, $\varphi(t)$ is the laser phase noise after arm locking, $\Phi_0(s)$ and $\Phi(s)$ are the Laplace transform of $\varphi_0(t)$ and $\varphi(t)$, respectively, $T$ is the round-trip time for the laser in the arm (for Taiji, $T \approx 20$ s), and $G(s)$ is the transfer function of the controller.

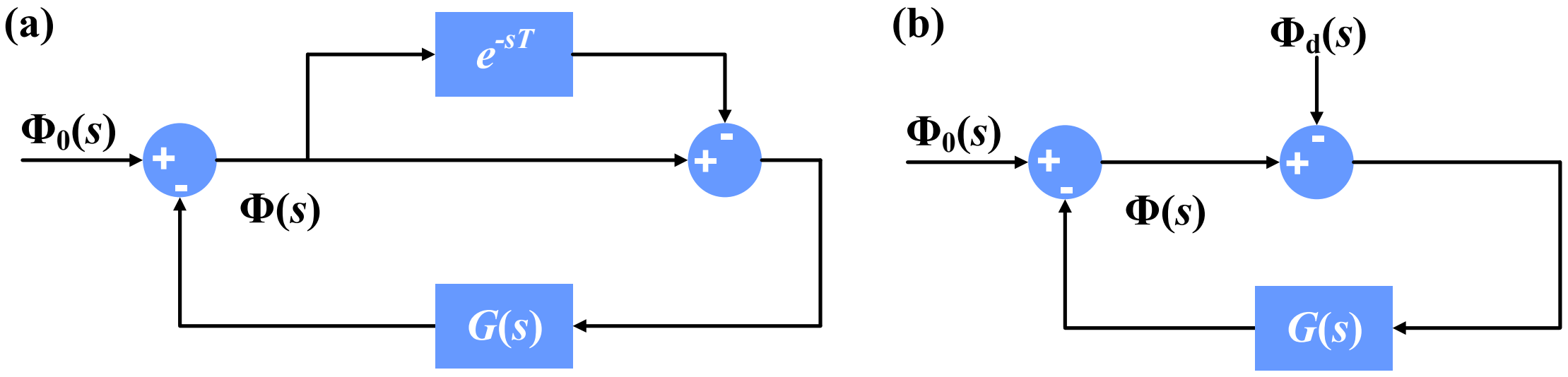

Figure 1. (a) diagram of single arm locking control system; (b) an equivalent block diagram of single arm locking.

Based on Fig. 1(a), the closed-loop transfer function can be easily obtained as:

$$H_{CL}(s) = \frac{\Phi(s)}{\Phi_0(s)} = \frac{1}{1+G(s)\left(1-e^{-sT}\right)} \tag{1}$$

$H_{\mathrm{CL}}(s)$ is a transcendental function due to the item $1\text{-}e^{-sT}$ in the denominator, making the transient analysis complicate. To solve this problem, the block $e^{-sT}$ can be replaced by an external input $\Phi_{\mathrm{d}}(s)$ in Fig. 1(b), where $\Phi_{\mathrm{d}}(s)= \Phi(s)e^{-sT}$. Then we have:

$$\Phi(s) = \frac{1}{1+G(s)}\Phi_0(s) + \frac{G(s)}{1+G(s)}\Phi_{\mathrm{d}}(s) \tag{2}$$

## A. Integral controller

To further simplify the model, let $G(s)=g/s$, where $g$ is the gain coefficient of the controller. Then Eq. (2) becomes:

$$\Phi(s) = \frac{s}{s+g}\Phi_0(s) + \frac{g}{s+g}\Phi_{\mathrm{d}}(s) \tag{3}$$

For stability analysis, we only need to investigate the step response of this system. Therefore, let $\varphi_0(t)=u(t)$, where $u(t)$ is the unit step function:

$$u(t) = \begin{cases} 0, & t \le 0^- \\ 1, & t \ge 0^+ \end{cases} \tag{4}$$

whose Laplace transform is $1/s$. Denote $p(t)$ as the corresponding time domain response of $\Phi(s)$ under input signal $u(t)$, then taking the inverse Laplace transform on both sides of Eq. (3), we have

$$p(t) = e^{-gt} + g\int_0^t e^{-g\tau} p(t-T-\tau)d\tau \tag{5}$$

Since $p(t)\equiv 0$ when $t<0$, if $0<t<T$, the integral item in Eq. (5) is zero, therefore:

$$p(t) = e^{-gt} \qquad 0<t<T \tag{6}$$

If $T\le t<2T$, substitute Eq. (6) into the integral item of Eq. (5), we obtain:

$$p(t) = e^{-gt} + g(t-T)e^{-g(t-T)} \qquad T \le t < 2T \tag{7}$$

Based on mathematical induction, for arbitrary integer $n\ge 0$, it can be easily derived:

$$p(t) = \sum_{k=0}^{n} \frac{g^k}{k!}(t-kT)^k e^{-g(t-kT)}, \qquad nT \le t < (n+1)T \tag{8}$$

Fig. 2 shows a typical $p(t)$ with $g$=10 and $T$=20 s. It can be seen that the initial step response decays pretty fast at the beginning, and the amplitude has been decreased to $4.54\times10^{-5}$ at 1 second (inset of Fig. 2). However, a new step response appears every ~20 seconds. The peaks of these step responses have a much slower damping speed than the initial step response. After 1000 seconds, the step response peak is still above 0.056.

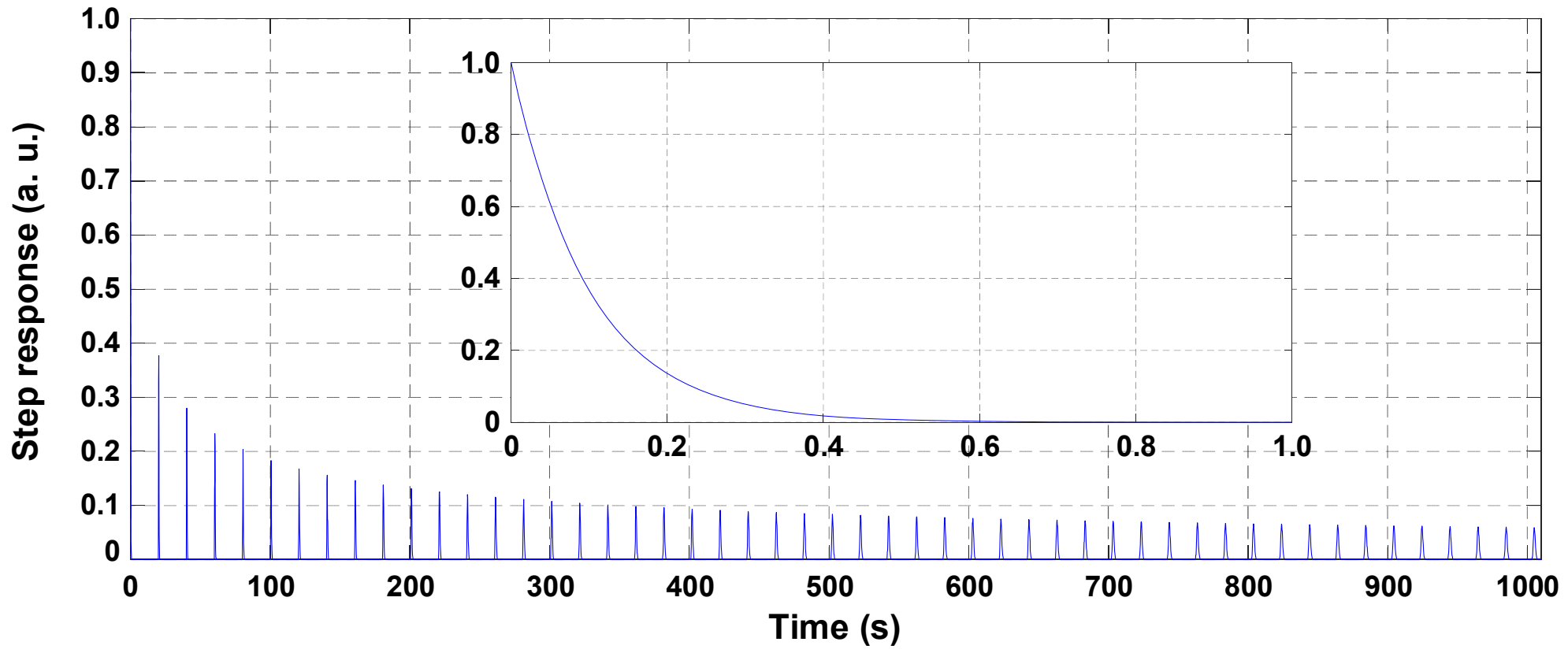

Figure 2. the step response of a single arm locking system with $G(s)$=10/$s$ and $T$=20 s.

In arm locking systems, usually $e^{-gT} << 1$ (e.g., for $g$=10, $T$=20 s, $e^{-gT}$ =1.384×10$^{-87}$). Therefore, all the items except the last one in the summation of Eq. (8) are negligible. Then the $n$-th step response can be approximated by:

$$p(t) \approx p_n(t-nT) = \frac{g^n}{n!}(t-nT)^n e^{-g(t-nT)}, \qquad nT \le t < (n+1)T \tag{9}$$

where

$$p_n(t) = \frac{g^n}{n!} t^n e^{-gt} \tag{10}$$

Using Eq. (10), the 1st, 2nd, 5th, 10th, 20th and 50th step response can be plotted in the same time interval, as shown in Fig. 3(a). Based on Eq. (3), the $n$-th step response is the output signal of the ($n$-1)-th step response after passing through the module with a transfer function of $g/(s+g)$. The inherent lag of this transfer function causes the peaks of those step responses in Fig. 3 (a) continuously shift to the right as $n$ increases.

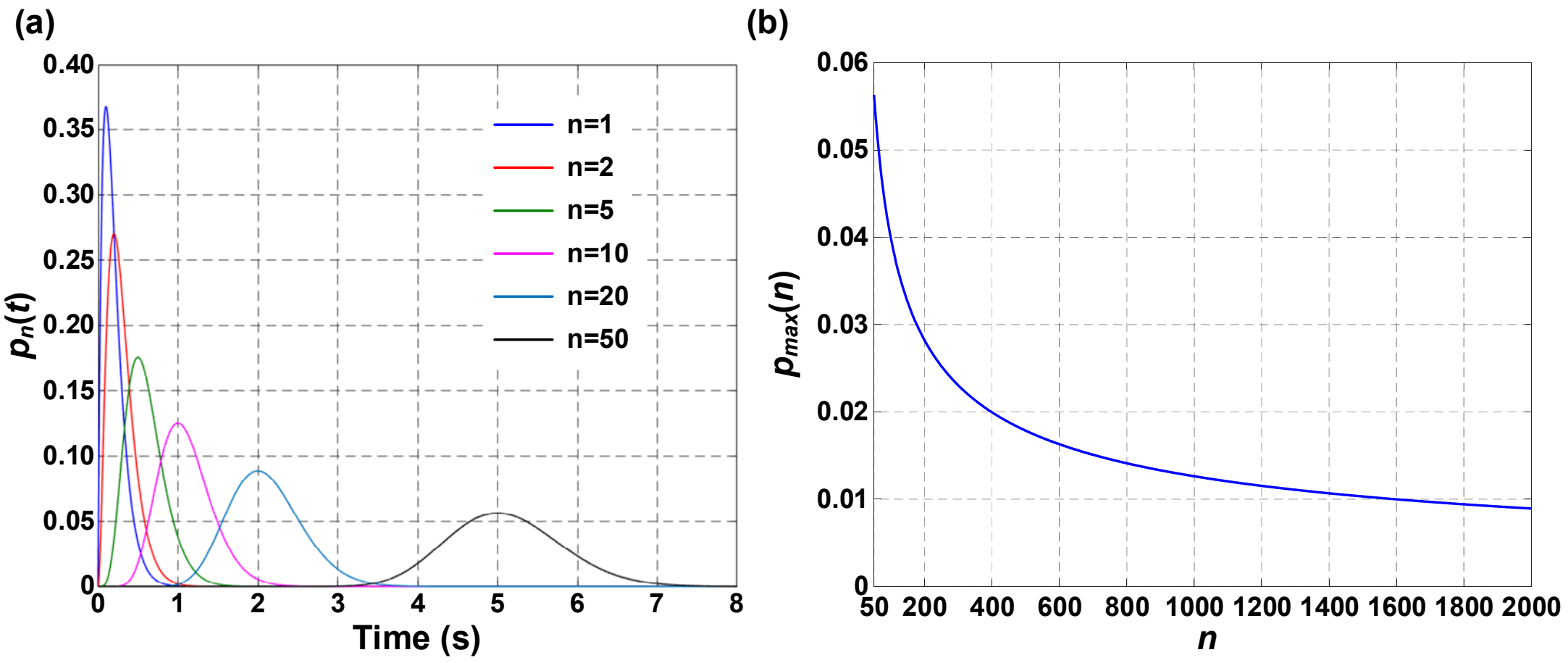

Figure 3. (a) $p_n(t)$ with different values of $n$; (b) $p_{max}(n)$ with different values of $n$.

By letting $p_n(t)'$ =0 ($p_n(t)'$ is the derivative of $p_n(t)$), the maximum value of $p_n(t)$ can be easily obtained as

$$\max\{p_n(t)\} = p_{\max}(n) = \frac{n^n}{n!} e^{-n} \tag{11}$$

Therefore, $p_{max}$ is independent with the gain coefficient $g$. $p_{max}$ is plotted as a function of $n$ in Fig. 3(b). Based on Stirling's approximation, for very large $n$,

$$p_{\max}(n) \approx \frac{1}{\sqrt{2\pi n}} \tag{12}$$

If we want the step response peak to be decreased to 10$^{-4}$, $n$ should be larger than 1.59×10$^7$, For Taiji this is equivalent to 10 years. So, using a simple controller $G(s)=g/s$, it is very difficult to damp the transient effect to be negligible.

### B. Stability criterion

Now we consider the transient response of a single arm locking system with a general controller given by:

$$G(s) = g\frac{B(s)}{A(s)} \tag{13}$$

where $g$ is the gain coefficient, $A(s)$ and $B(s)$ are monic polynomials of $s$ with real coefficients. Substitute Eq. (13) into Eq. (2),

$$\Phi(s)=\frac{A(s)}{A(s)+g\times B(s)}\Phi_0(s)+\frac{g\times B(s)}{A(s)+g\times B(s)}\mathscr{L}\left\{\varphi(t-T)\right\} \tag{14}$$

Since $\varphi(t)\equiv 0$ when $t<0$, if $0<t<T$, we have

$$\varphi(t)=\mathscr{L}^{-1}\left\{\frac{A(s)}{A(s)+g\times B(s)}\right\}\otimes\varphi_0(t) \quad 0<t<T \tag{15}$$

For simplicity, Eq. (15) can be rewritten in $s$ domain as

$$\Phi(s)=\frac{A(s)}{A(s)+g\times B(s)}\Phi_0(s) \quad 0<t<T \tag{16}$$

(16) can be understood as: the inverse Laplace transform of this equation holds when $0<t<T$. In the following of this paper, we will use this simplified expression without explanation. If $T\le t<2T$, substitute Eq. (16) into Eq. (14), it can be obtained:

$$\Phi(s)=\frac{A(s)}{A(s)+g\times B(s)}\Phi_0(s)+\frac{g\times B(s)}{A(s)+g\times B(s)}\frac{A(s)}{A(s)+g\times B(s)}\Phi_0(s)e^{-sT} \quad T\le t<2T \tag{17}$$

Based on mathematical induction, for arbitrary integer $n\ge 0$, it can be derived:

$$\Phi(s)=\left\{\frac{A(s)}{A(s)+g\times B(s)}\sum_{k=0}^{n}\left[\frac{g\times B(s)}{A(s)+g\times B(s)}\right]^k e^{-skT}\right\}\Phi_0(s), \quad nT\le t<(n+1)T \tag{18}$$

So, the closed-loop transfer function is

$$H_{CL}(s)=\frac{A(s)}{A(s)+g\times B(s)}\sum_{k=0}^{n}\left[\frac{g\times B(s)}{A(s)+g\times B(s)}\right]^k e^{-skT}, \quad nT\le t<(n+1)T \tag{19}$$

Defining

$$H_0(s)=\frac{A(s)}{A(s)+g\times B(s)} \tag{20}$$

$$H(s)=\frac{g\times B(s)}{A(s)+g\times B(s)} \tag{21}$$

Then Eq. (19) can be written as

$$H_{CL}(s)=\sum_{k=0}^{n}H_0H(s)^k e^{-skT}, \quad nT\le t<(n+1)T \tag{22}$$

Suppose:

$$A(s)+g\times B(s)=(s-p_1)^{q_1}(s-p_2)^{q_2}\cdots(s-p_r)^{q_r} \tag{23}$$

where $p_1$, …, $p_r$ are poles of $H_0(s)$ and $H(s)$, and $q_1$, …, $q_r$ are natural numbers. Based on [33], $H_0(s)$ and $H(s)$ are stable if all $p_i$ ($i$=1, …, $r$) have negative real parts (i.e., they are all in the left-hand $s$-plane) and are unstable otherwise.

To guarantee $H_{CL}(s)$ is stable, all the items on the right side of Eq. (22) should be stable. Therefore, for arbitrary integer $n\ge 1$, $H_0(s)H(s)^n$ needs to be stable. Suppose $h_0(t)$ and $h_n(t)$ are the inverse Laplace transform of $H_0(s)$ and $H_0(s)H(s)^n$, respectively. Then $h_n(t)$ can be regarded as the output of the input signal $h_0(t)$ after passing through the filter $H(s)$ $n$ times.

Let $s=j\omega$, where $\omega$ is the angular frequency. If $\exists$ frequency interval $[\omega_1, \omega_2]$, and $|H(j\omega)|>1$ when $\omega\in[\omega_1, \omega_2]$, then the frequency components of $h_0(t)$ within $[\omega_1, \omega_2]$ will be amplified by $H(s)$. when $n\to\infty$, the amplitude of these frequency components will go to infinity, thus cause the instability of $H_{CL}(s)$.

On the other hand, if for arbitrary frequency $\omega$, $|H(j\omega)|\leq 1$, then based on Eq. (21), we have $|H(j\omega)|<1$ when $\omega\to\infty$, thus $|H(j\omega)|$ cannot always be 1. Suppose $[\omega_1, \omega_2]$ is the longest frequency interval in which $|H(j\omega)|=1$ is satisfied. Then if $\omega_1\neq\omega_2$, the derivative of $|H(j\omega)|^2$ will be discontinuous at $\omega_1^+$ and $\omega_1^-$ (as well as $\omega_2^+$ and $\omega_2^-$), which is not possible since both the numerator and denominator of $|H(j\omega)|^2$ are polynomial of $\omega$. Therefore, $|H(j\omega)|=1$ can only happen at a few isolated frequencies. When $h_0(t)$ passes through $H(s)$, the amplitude of those isolated frequency components of $h_0(t)$ will keep unchanged, while all the other frequency components will be attenuated. When $n\to\infty$, all the other frequency components will be attenuated to zero, only those isolated frequency components of $h_0(t)$ are left in $h_n(t)$. Due to the stability of $H_0(s)$, we have $h_0(t)=0$ when $t\to\infty$, which means any single frequency component contribution to $h_0(t)$ is negligible (i.e., $h_0(t)$ cannot be a DC signal, a sinusoidal signal or a periodical signal). Therefore, $h_n(t)$ is also negligible when $n\to\infty$, and then $H_{CL}(s)$ is stable.

Based on the discussions above, the stability criterion for a single arm locking controller can be summarized as:

I. All the poles of $H(s)$ have negative real parts:

$$\mathrm{Re}(p_i)<0,\quad i=1,2\cdots,r \tag{24A}$$

II. The amplitude response of $H(s)$ is always not higher than 1:

$$\max_{\omega\in[-\infty,+\infty]}|H(j\omega)|\leq 1 \tag{24B}$$

If criterion (24A) and (24B) are satisfied, Eq. (22) is equivalent to Eq. (1) when $n\to\infty$, which can be easily justified by the summation formula of geometric series.

In practice, different design rules can be derived from (24A) and (24B), depending on the specific forms of $A(s)$ and $B(s)$. Let

$$A(s)=s^m+a_{m-1}s^{m-1}+\cdots a_0 \tag{25}$$

$$B(s)=s^l+b_{l-1}s^{l-1}+\cdots b_0 \tag{26}$$

To guarantee $|H(j0)|\leq 1$, we have:

$$\left|\frac{gb_0}{a_0+gb_0}\right|\leq 1 \tag{27}$$

Based on Eq. (23), $H(s)$ can be written in the form of partial-fraction expansion as

$$H(s)=C_0+\frac{C_{11}}{s-p_1}+\frac{C_{12}}{(s-p_1)^2}+\ldots\frac{C_{1q_1}}{(s-p_1)^{q_1}}+\frac{C_{21}}{s-p_2}+\ldots\frac{C_{2q_2}}{(s-p_2)^{q_2}}+\ldots\frac{C_{rq_r}}{(s-p_r)^{q_r}} \tag{28}$$

where $C_0$, $C_{ik}$ ($i$=1, 2, …, $r$, $k$=1, 2, …, $q_i$) are constant coefficients. Let $s$=$j$0 in Eq. (28), Eq. (27) is equivalent to

$$\left|C_0+\frac{C_{11}}{-p_1}+\frac{C_{12}}{(-p_1)^2}+\ldots\frac{C_{1q_1}}{(-p_1)^{q_1}}+\frac{C_{21}}{-p_2}+\ldots\frac{C_{2q_2}}{(-p_2)^{q_2}}+\ldots\frac{C_{rq_r}}{(-p_r)^{q_r}}\right|\leq 1 \tag{29}$$

Generally, (29) is only a necessary but not sufficient condition for the stability of $H_{CL}(s)$, but in certain cases it can also be sufficient. Based on Eq. (28), for arbitrary frequency $\omega$,

$$\begin{aligned}|H(j\omega)|&=\left|C_0+\frac{C_{11}}{j\omega-p_1}+\frac{C_{12}}{(j\omega-p_1)^2}+\ldots\frac{C_{1q_1}}{(j\omega-p_1)^{q_1}}+\frac{C_{21}}{j\omega-p_2}+\ldots\frac{C_{2q_2}}{(j\omega-p_2)^{q_2}}+\ldots\frac{C_{rq_r}}{(j\omega-p_r)^{q_r}}\right|\\&\leq|C_0|+\left|\frac{C_{11}}{j\omega-p_1}\right|+\left|\frac{C_{12}}{(j\omega-p_1)^2}\right|+\ldots\left|\frac{C_{1q_1}}{(j\omega-p_1)^{q_1}}\right|+\left|\frac{C_{21}}{j\omega-p_2}\right|+\ldots\left|\frac{C_{2q_2}}{(j\omega-p_2)^{q_2}}\right|+\ldots\left|\frac{C_{rq_r}}{(j\omega-p_r)^{q_r}}\right|\end{aligned} \tag{30}$$

If $p_i$ ($i$=1, 2, …, $r$) are negative real numbers, and $C_0\geq 0$, $C_{ik}\geq 0$ ($i$=1, 2, …, $r$, $k$=1, 2, …, $q_i$), (30) can be simplified as

$$|H(j\omega)| \le C_0 + \frac{C_{11}}{-p_1} + \frac{C_{12}}{(-p_1)^2} + \dots \frac{C_{1q_1}}{(-p_1)^{q_1}} + \frac{C_{21}}{-p_2} + \dots \frac{C_{2q_2}}{(-p_2)^{q_2}} + \dots \frac{C_{rq_r}}{(-p_r)^{q_r}} \quad (31)$$

Based on (31) and (24B), condition (29) can be used as a stability criterion for $H_{CL}(s)$ under the assumption above. Condition (29) is a practical design rule if we start the controller design from the partial-fraction expansion form in Eq. (28).

### C. Proportional integral controller

Suppose $G(s)$ is a proportional integral controller given by

$$G(s) = g\frac{s+a}{s} \quad (32)$$

Then

$$H(s) = \frac{gs+ga}{(1+g)s+ga} \quad (33)$$

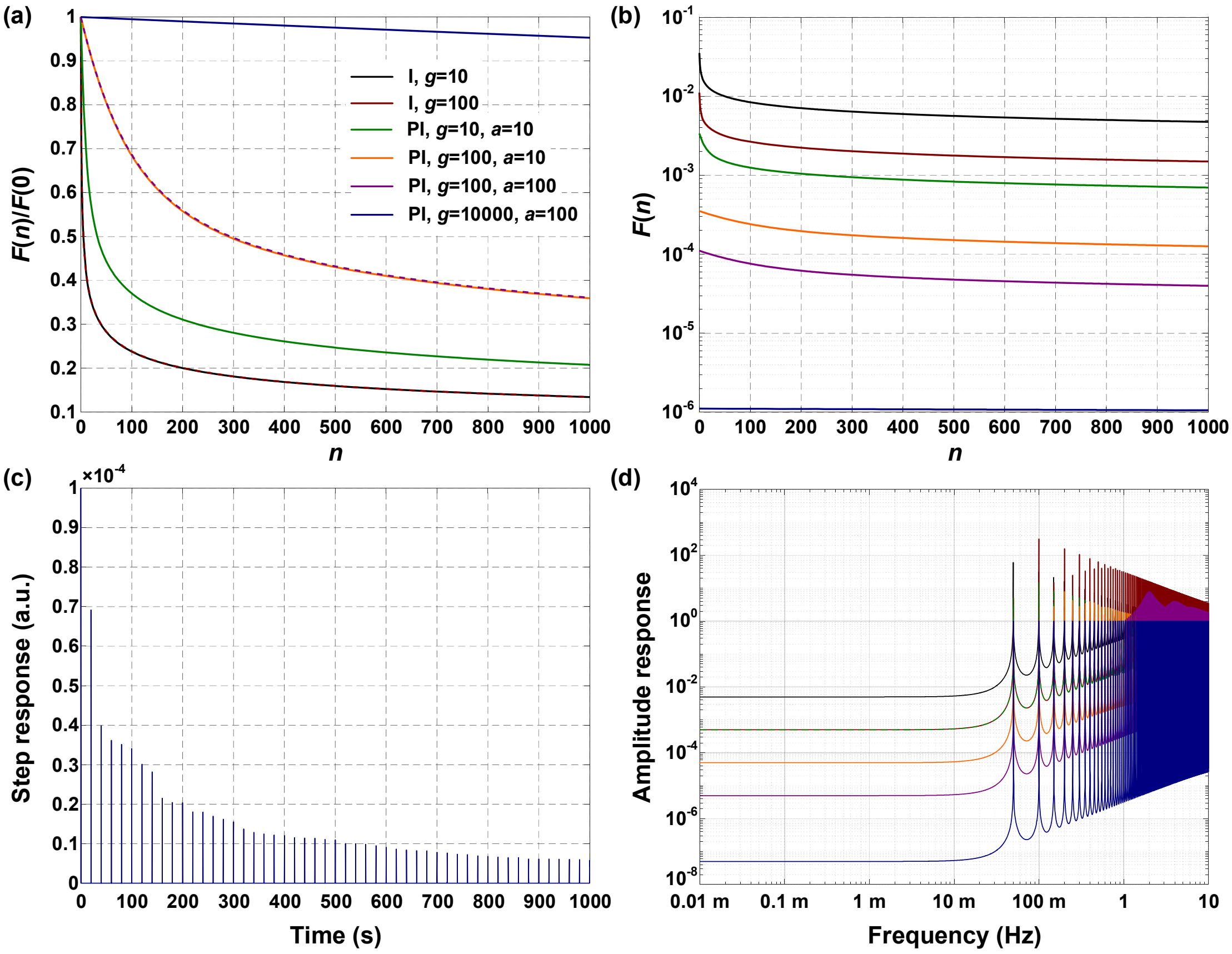

Figure 4. (a) $F(n)/F(0)$ for six arm locking controllers; (b) $F(n)$ for six arm locking controllers; (c) the step response of a PI controller with $g$=10000 and $a$=100; (d) the amplitude response of $H_{CL}(s)=1/[1+G(s)(1-e^{sT})]$ for six arm locking controllers; four figures share the same legend in (a), for visibility, some curves are changed to dashed lines when they coincide with other curves.

It can be easily verified that, as long as $g>0$, $a>0$, the stability criterion (24A) and (24B) are automatically satisfied. Since there is no limitation on $g$ and $a$, large numbers can be adopted to achieve high laser noise suppression ratio.

In an arm locking system, usually $h_n(t)$ decays to almost zero within the time of $T$, therefore, in Eq. (22), $H_{CL}(s)$ is mainly determined by the last item in the summation. Then based on Parseval's theorem, the RMS value of the step response of $H_{CL}(s)$ during the period of $[nT, (n+1)T]$ can be defined as

$$F(n)=\left[\frac{1}{T}\int_{-\infty}^{+\infty}\left|\frac{1}{s}H_0(j\omega)H(j\omega)^n\right|^2 d\omega\right]^{\frac{1}{2}} \tag{34}$$

We can use $F(n)$ to investigate the decay rate of the periodically appeared transient responses.

The relative value $F(n)/F(0)$ of two integral (I) controllers and four proportional integral (PI) controllers are compared and the results are shown in Fig. 4 (a). It can be seen that I controller gives the higher decay rate, which is almost independent with $g$. While the decay rate of PI controller is independent with $a$, but slows down as $g$ increases. Although the transient responses of PI controllers decay slower, due to the high suppression ratio provided by $H_0$, the initial transient response of a PI controller is much smaller than that of an I controller with the same $g$ value, as indicated by the $F(n)$ values in Fig. 4 (b). With $g$=10000 and $a$=100, the initial transient response peak of a PI controller is only $1\times10^{-4}$ (Fig. 4(c)) and can be decreased to below $1\times10^{-5}$ after 1000 s. Utilizing its good transient response characteristics and high noise suppression ratio in the frequency domain (Fig. 4(d)), the high gain PI controller can serve as a candidate for arm locking system.

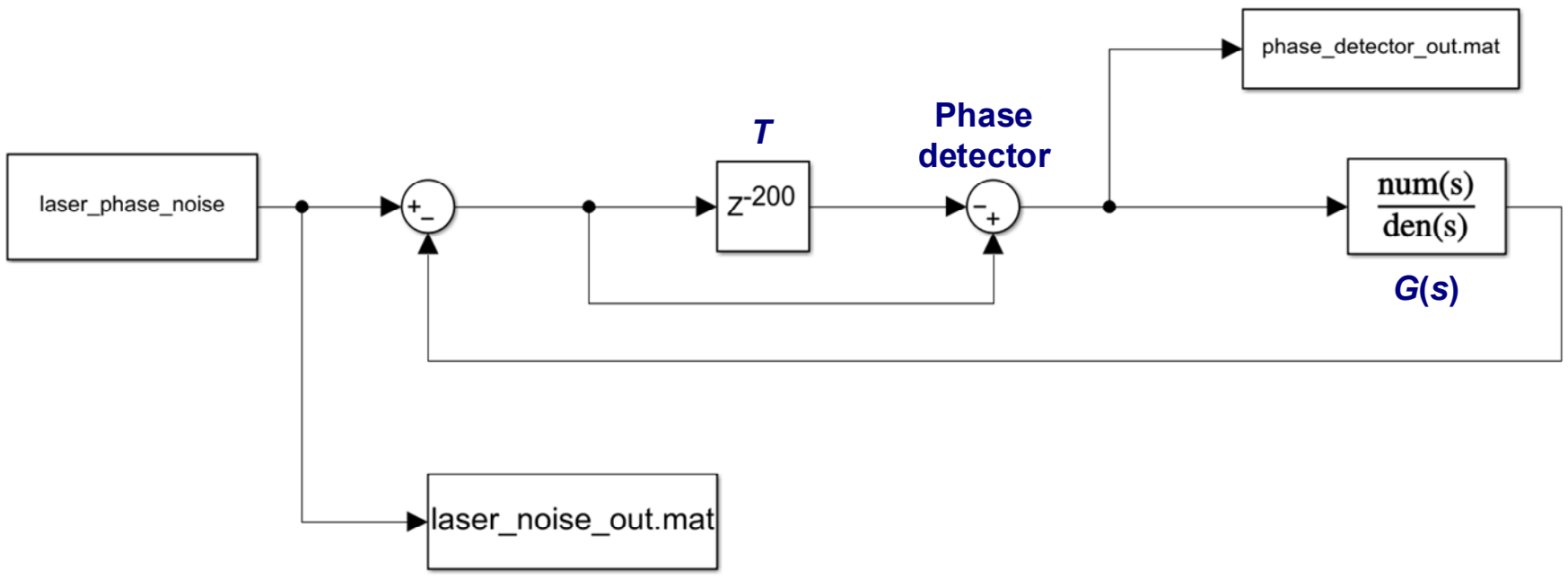

Figure 5. Simulink diagram of single arm locking with PI controller, $T$=20 s, $G(s)$=10000($s$+100)/$s$.

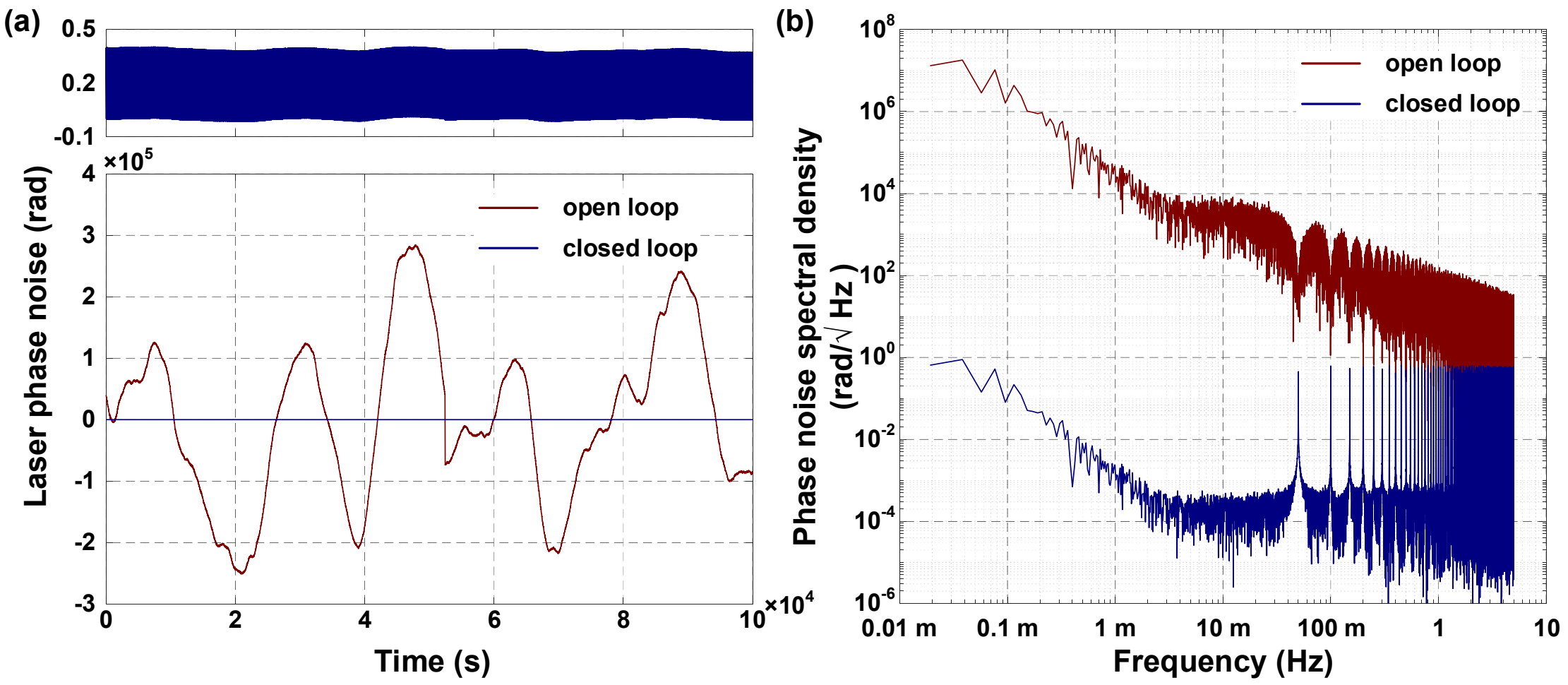

Figure 6. Single arm locking simulation results using PI controller. (a) laser phase noise after phase detector in time domain; (b) laser phase noise spectral density.

To further evaluate the performance of a high gain PI controller in arm locking, a time domain MATLAB/Simulink simulation is performed. The diagram is depicted in Fig. 5. The reason for using a discrete delay element is to guarantee that in the first 20 s (i.e., $T$) the delay element can output the real laser phase noise, which agrees with the physical experiments, while continuous delay element can only output zeros in the first 20 s due to software limitation. Besides, since in real arm locking system, the physically measurable signal is the phase difference after phase detector, in the simulation, we also choose the phase detector's output to evaluate the locking performance.

The simulation is executed for $10^5$ seconds. The phase detector's output signals for open loop and closed loop are compared in Fig. 6 (a). The RMS laser phase noise is reduced from $1.31\times10^5$ rad (open loop) to 0.0282 rad (close loop). In the upper part of Fig. 6 (a), the closed-loop results are zoomed in to show the details. Due to the good transient response characteristics of the high gain PI controller, the closed-loop results enter the steady state almost as soon as the arm locking is started. The open-loop and closed-loop phase noise spectral densities are presented in Fig. 6 (b). Except for those frequencies around the dead zones of the interferometer ($n/T$, i.e., $0.05\times n$ Hz), the laser phase noise can be suppressed by more than 5 orders within the full science band (0.1 mHz to 1 Hz).

### D. High-order controllers

Now we consider high-order controllers when $m\geq2$ or $l\geq2$ in Eq. (25) and (26). If criterion (24A) and (24B) are satisfied, the stable form of the closed-loop transfer function is

$$H_{CL}(s)=\frac{A(s)}{A(s)+g\times B(s)(1-e^{-sT})} \tag{35}$$

The low frequency limit of Eq. (35) is

$$\lim_{s\to0}H_{CL}(s)=\frac{A(s)}{A(s)+g\times B(s)\times sT} \tag{36}$$

If $a_0\neq0$, this limit is equal to 1. Therefore, to achieve high noise suppression ratio at low frequencies, $a_0$ must be zero. If $a_i=0$ ($i$=0, 1, …, $k$-1), $a_k\neq0$,

$$\lim_{s\to0}H_{CL}(s)=\frac{a_k}{a_k+gTb_0} \tag{37}$$

High noise suppression ratio requires:

$$gTb_0 \gg a_k \tag{38}$$

Suppose $m$=2 and $l$=2, $G(s)$ is given by

$$G(s)=g\frac{s^2+bs+c}{s^2+as} \tag{39}$$

Then

$$H(s)=\frac{gs^2+gbs+gc}{(1+g)s^2+(a+gb)s+gc} \tag{40}$$

To meet criterion (24B), the following condition needs to be satisfied:

$$\left|-g\omega^2+gbj\omega+gc\right|^2\leq\left|-(1+g)\omega^2+(a+gb)j\omega+gc\right|^2 \tag{41}$$

(41) can be simplified as

$$a^2+2gab\geq2gc \tag{42}$$

On the other hand, to guarantee high noise suppression ratio at low frequencies, we need

$$gTc \gg a \tag{43}$$

We choose three sets of parameters: {$a$, $b$, $c$, $g$} = {1,10000, 10000, 100}, {1,1000, 1000, 1000} and {1, 100, 100, 10000}, all of which can meet (42), (43) and criterion (24A) simultaneously. The transient characteristics of these three controllers are also analyzed using $F(n)$ function and compared with that of a PI controller with $g$=10000 and $a$=100. Although the 2nd-order controllers can give faster decay rate (Fig. 7(a)), their initial $F(n)$ values are much higher than that of the PI controller (Fig. 7(b)), which is attributed to the lower noise suppression ratio of their $H_0$. With the same gain value $g$=10000, the step response peaks of the 2nd-order controller also decay slower than that of the PI controller (Fig. 7(c) and Fig. 4(c)). Therefore, although the four controllers have almost the same closed-loop amplitude response (Fig. 7(d)), there real performances in an arm locking system could be pretty different.

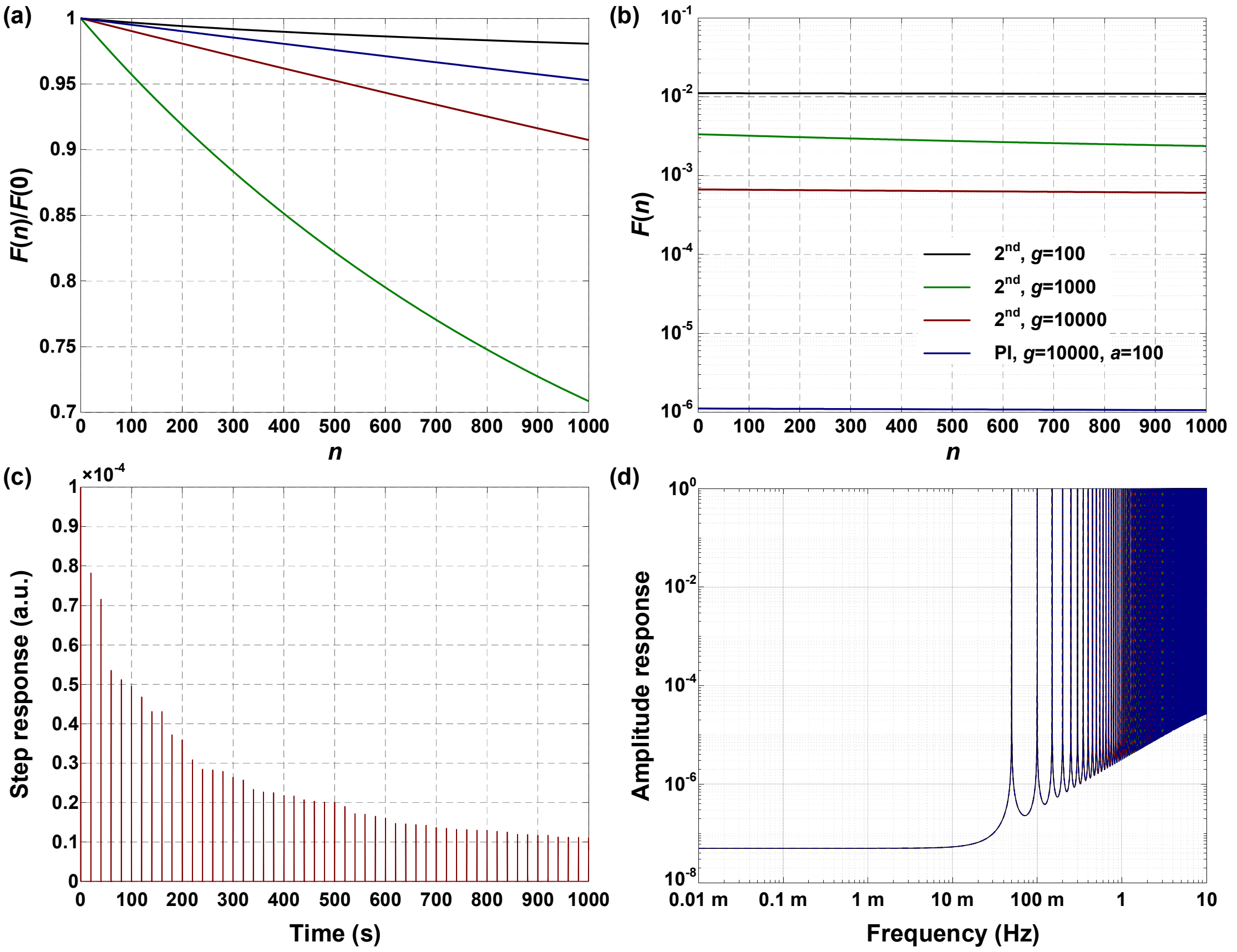

Figure 7. (a) $F(n)/F(0)$ for four arm locking controllers; (b) $F(n)$ for four arm locking controllers; (c) the step response of 2$^{nd}$-order controller with $a$=1, $b$=100, $c$=100 and $g$=10000; (d) the amplitude response of $H_{CL}(s)=1/[1+G(s)(1-e^{sT})]$ for four arm locking controllers; four figures share the same legend in (b), for visibility, some curves are changed to dashed lines when they coincide with other curves.

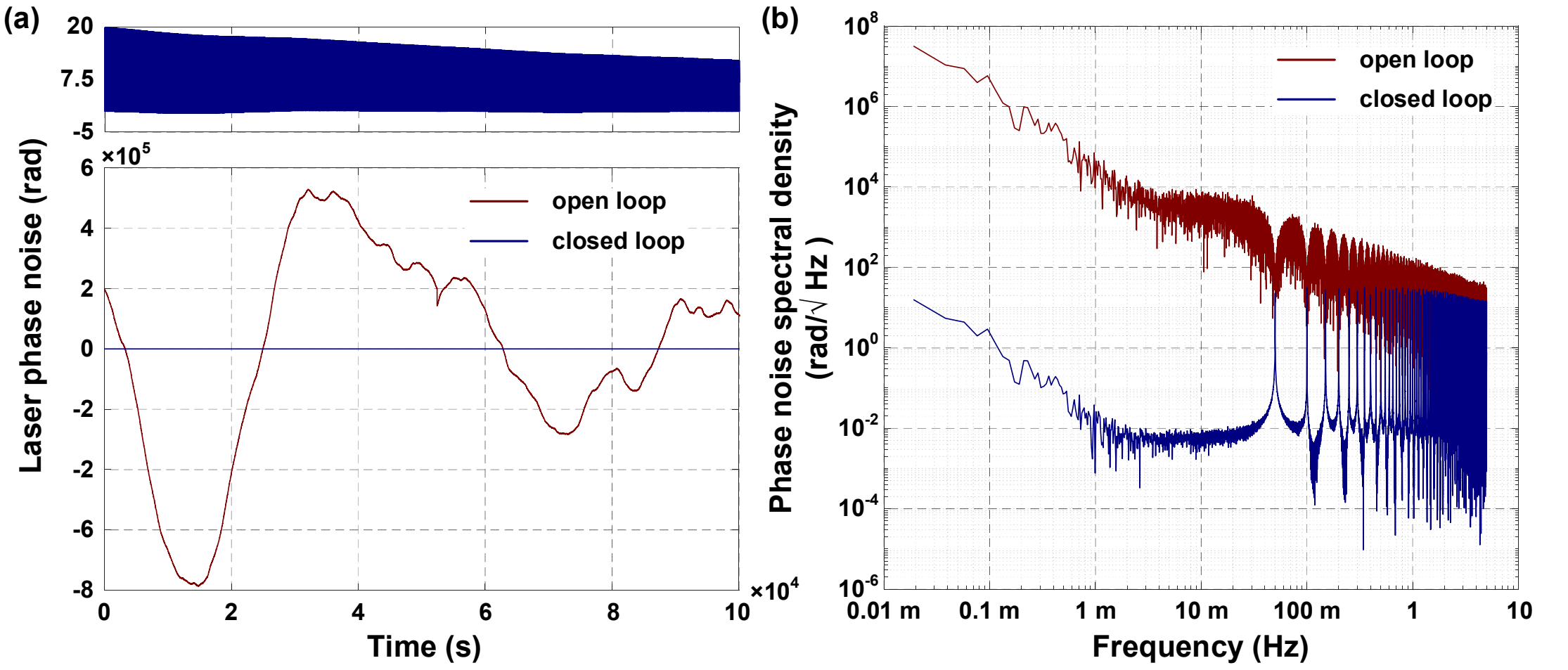

Figure 8. Single arm locking simulation results using 2$^{nd}$-order controller ($a$=1, $b$=100, $c$=100, $g$=10000). (a) laser phase noise after phase detector in time domain; (b) laser phase noise spectral density.

For example, the 2$^{nd}$-order controller with $g$=10000 is simulated in time domain using the same setup of Fig. 5. During the whole $10^5$ s simulation time, the peak-to-peak closed-loop phase noise value continuously decreases (Fig. 8(a), upper part), which means the system has not entered the steady state yet. This imperfect transient response results in a little worse noise suppression ratio in Fig. 8(b), compared with the results in Fig. 6(b).

Generally, any high-order controllers can be designed by following the similar procedures like (39) to (43). Based on our experiences so far, large value of $g$ is necessary to reduce the initial value of $F(n)$, and as a result to improve the transient response performance. On the other hand, the higher order the controller is, the more difficult to meet criterion (24B) using large $g$ values. With same $g$ value, PI controller can provide the better time domain performance than other high-order controllers we have tested.

## 3. Dual arm locking

The inherent dead zones around the frequencies of $n/T$ limit the laser phase noise suppression performance of single arm locking. To resolve this issue, dual arm locking was proposed [13] to push the first dead zone frequency, $1/T$, out of the science band. In this section, we will discuss the transient stability of dual arm locking controller.

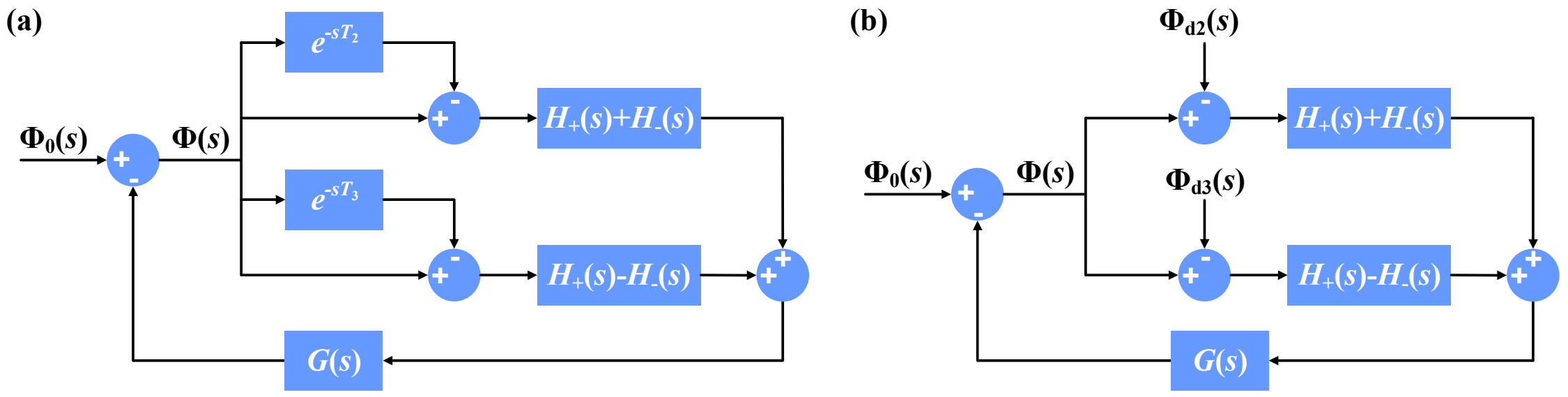

Figure 9. (a) diagram of dual arm locking control system; (b) an equivalent block diagram of dual arm locking.

The diagram of dual arm locking control system is given in Fig. 9(a). $T_2$ and $T_3$ ($T_2>T_3$) are the round-trip time between spacecraft 1 and 2, and spacecraft 1 and 3, respectively. $G(s)$ is the controller defined by the general form in Eq. (13). $H_+(s)$ and $H_-(s)$ are dual arm locking sensors. Based on this diagram, the closed-loop transfer function of dual arm locking is:

$$H_{CL}(s)=\frac{\Phi(s)}{\Phi_0(s)}=\frac{1}{1+G(s)(1-e^{-sT_2})(H_+(s)+H_-(s))+G(s)(1-e^{-sT_3})(H_+(s)-H_-(s))} \tag{44}$$

To make it easier to analyze the transient response, the two delay elements can be replaced by two external input signals $\Phi_{d2}(s)$ and $\Phi_{d3}(s)$ in Fig. 9(b), where

$$\Phi_{d2}(s)=\Phi(s)e^{-sT_2} \tag{45}$$

$$\Phi_{d3}(s)=\Phi(s)e^{-sT_3} \tag{46}$$

Based on the diagram of Fig. 9(b), we have

$$\Phi(s)=\frac{1}{1+2H_+G}\Phi_0(s)+\frac{(H_++H_-)G}{1+2H_+G}\Phi_{d2}(s)+\frac{(H_+-H_-)G}{1+2H_+G}\Phi_{d3}(s) \tag{47}$$

Using similar derivations as Eq. (15)-(18), it can be obtained:

$$\Phi(s)=\left\{\frac{1}{1+2H_+G}\sum_{k=0}^{n}\left[\frac{H_++H_-}{1+2H_+G}Ge^{-sT_2}+\frac{H_+-H_-}{1+2H_+G}Ge^{-sT_3}\right]^k\right\}\Phi_0(s),\quad nT_2\le t<mT_2+lT_3 \tag{48}$$

where $mT_2+lT_3$ is the minimum value that is larger than $nT_2$, for arbitrary non-negative integers $m$ and $l$.

Defining

$$\overline{\tau}=\frac{T_2+T_3}{2} \tag{49}$$

$$\Delta\tau = \frac{T_2 - T_3}{2} \tag{50}$$

$$H_{0,\text{dual}}(s) = \frac{1}{1 + 2H_+(s)G(s)} \tag{51}$$

$$H_{\text{dual}}(s) = \frac{H_+ + H_-}{1 + 2H_+ G} G e^{-s\Delta\tau} + \frac{H_+ - H_-}{1 + 2H_+ G} G e^{s\Delta\tau} \tag{52}$$

Based on (49)-(52), the closed-loop transfer function becomes

$$H_{CL}(s) = H_{0,\text{dual}} \sum_{k=0}^{n} H_{\text{dual}}^{k} e^{-sk\bar{\tau}}, \quad nT_2 \le t < mT_2 + lT_3 \tag{53}$$

Similar to the analysis of single arm locking, the stability criterion for a dual arm locking controller can be summarized as:

I. All the poles of $H_{\text{dual}}(s)$ have negative real parts:

$$\text{Re}(p_{i,\text{dual}}) < 0, \quad i = 1, 2 \cdots, r \tag{54A}$$

II. The amplitude response of $H_{\text{dual}}(s)$ is always not higher than 1:

$$\max_{\omega \in [-\infty, +\infty]} \left| H_{\text{dual}}(j\omega) \right| \le 1 \tag{54B}$$

As an example, let $\bar{\tau} = 20$ s, $\Delta\tau$=0.1 s and

$$H_+(s) = 1 \tag{55}$$

$$H_(s) = \frac{1}{s\Delta\tau} \frac{2\pi f_0}{s + 2\pi f_0} \tag{56}$$

If a high gain PI controller with $g$=10000 and $a$=100 is used for $G(s)$, criterion (54B) requires $f_0$<0.49 Hz.

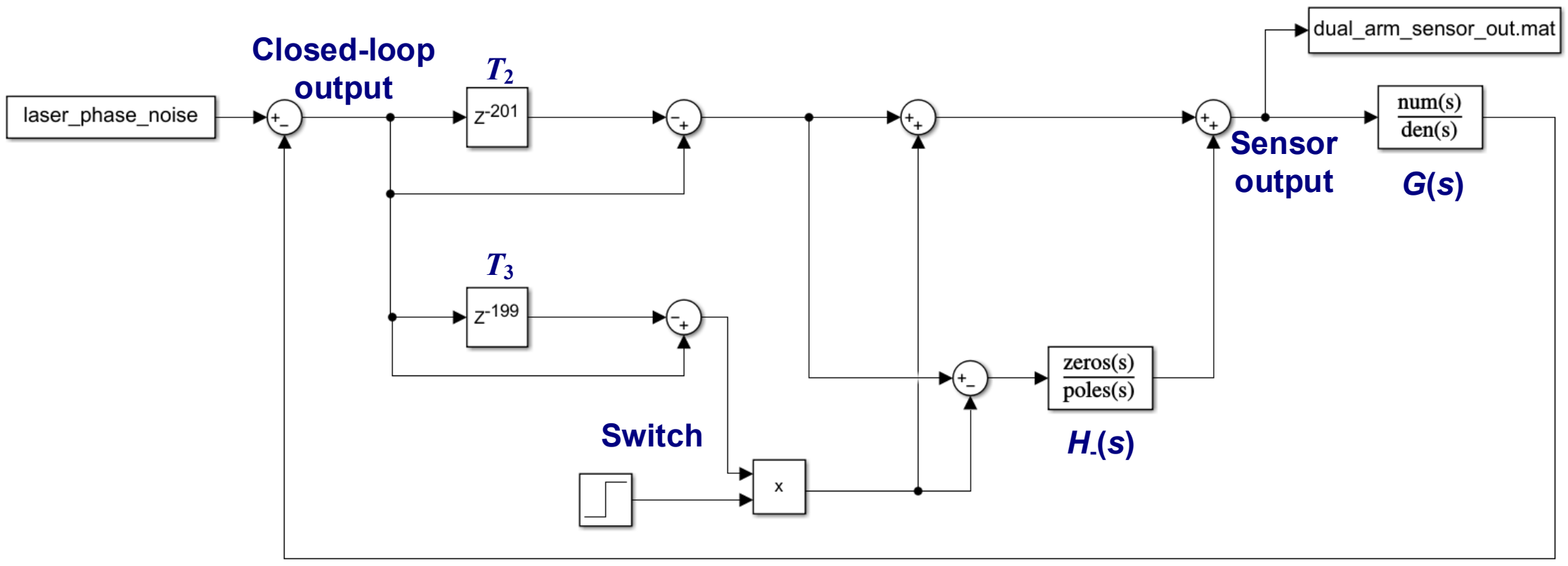

Figure 10. Simulink diagram of dual arm locking with a PI controller, $\Delta\tau$=0.1 s, $f_0$= 0.48 Hz, $G(s)$=10000($s$+100)/$s$.

Dual arm locking with a PI controller is also simulated in time domain by Simulink using the diagram of Fig. 10. A switch consists of a step function and a multiply function is added to turn on the $T_3$ arm after 25 s, so as to make the locking start smoothly. The laser phase noise at the sensor output (the physically measurable signal) for open loop and closed loop are compared in Fig. 11 (a). The RMS laser phase noise is reduced from 4.87×10$^8$ rad (open loop) to 0.246 rad (close loop). The transient response peaks continue to decay throughout the entire simulation time, from ~5.5 rad at the beginning (Fig. 11 (a) inset) to about ~0.04 rad at 10$^5$ s (Fig. 11 (a) upper part). The spectral density of the phase noise data from 5×10$^4$ s to 10$^5$ s are calculated and shown in Fig. 11(b). The laser phase noise is suppressed by more than 8 orders of magnitude at 0.1 mHz, and between 4 and 5 orders around 1 Hz. The $n / \bar{\tau}$ peaks have been perfectly eliminated by the dual arm locking sensors.

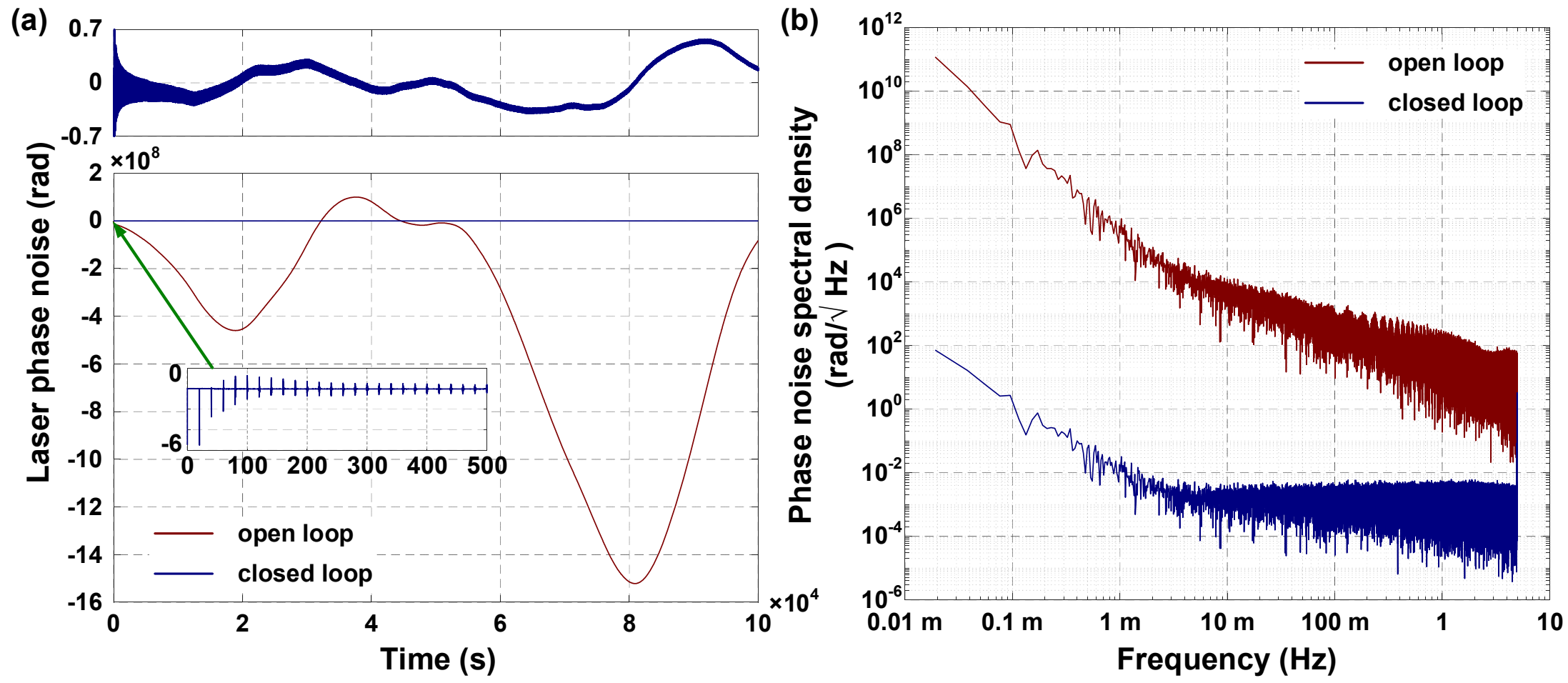

Figure 11. Dual arm locking simulation results using a PI controller. (a) laser phase noise in time domain; (b) laser phase noise spectral density.

## 4. Modified-dual arm locking

Our stability criterion can also be applied to modified-dual arm locking, which is proposed to handle the Doppler frequency pulling problem [14]. In modified-dual arm locking, $H_+(s)$ and $H_-(s)$ in Fig. 9(a) are replaced by

$$H_{M+}(s) = \left[F_C(s) + F_D(s)\right] H_+(s) \tag{57}$$

$$H_{M-}(s) = F_D(s) H_-(s) \tag{58}$$

where $F_C(s)$ and $F_D(s)$ are two specifically designed filters, $F_C(s)$ dominates at low frequencies, while $F_D(s)$ dominates at high frequencies. Then the closed-loop transfer function in Eq. (53) needs to be modified to

$$H_{CL}(s) = H_{0,M} \sum_{k=0}^{n} H_M^k e^{-sk\bar{\tau}}, \quad nT_2 \le t < mT_2 + lT_3 \tag{59}$$

where

$$H_{0,M}(s) = \frac{1}{1 + 2H_{M+}(s)G(s)} \tag{60}$$

$$H_M(s) = \frac{H_{M+} + H_{M-}}{1 + 2H_{M+}G} G e^{-s\Delta\tau} + \frac{H_{M+} - H_{M-}}{1 + 2H_{M+}G} G e^{s\Delta\tau} \tag{61}$$

The stability criterion for this modified-dual arm locking controller is:

I. All the poles of $H_M(s)$ have negative real parts:

$$\mathrm{Re}(p_{i,M}) < 0, \quad i = 1, 2 \cdots, r \tag{62A}$$

II. The amplitude response of $H_M(s)$ is always not higher than 1:

$$\max_{\omega \in [-\infty, +\infty]} \left| H_M(j\omega) \right| \le 1 \tag{62B}$$

It is quite challenging to design $F_C(s)$ and $F_D(s)$ that can satisfy criterion (62A) and (62B) simultaneously. Here we only provide a note during the design. In an arm locking system, usually $G$>>1, thus the zeros of $F_C(s) + F_D(s) = 0$ should be very close to the poles of $H_M(s)$. To meet (62A), we need to first guarantee that all zeros of $F_C(s) + F_D(s) = 0$ have negative real parts. Assume $F_C(s)$ is an $n$-order low pass filter (LPF) and $F_D(s)$ is an $m$-order high pass filter (HPF):

$$F_C(s) = \prod_{k=1}^{n} \frac{g_{Lk}}{s + p_{Lk}} \tag{63}$$

$$F_D(s)=\prod_{k=1}^{m}\frac{g_{Hk}s}{s+p_{Hk}} \tag{64}$$

Then $F_C(s)+F_D(s)=0$ is equivalent to

$$\prod_{k=1}^{n}g_{Lk}\prod_{k=1}^{m}(s+p_{Hk})+s^m\prod_{k=1}^{m}g_{Hk}\prod_{k=1}^{n}(s+p_{Lk})=0 \tag{65}$$

If $n\geq 2$, $s^{m+n-1}$ item is missing in Eq. (65), based on Theorem 1 of Appendix, some roots of Eq. (65) would not have negative real parts. Therefore, to meet criterion (62A) with a high gain controller, $F_C(s)$ can only be a first order LPF:

$$F_C(s)=\frac{g_a}{s+2\pi f_a} \tag{66}$$

However, our numerical calculation shows that it is difficult to meet criterion (62B) if $F_C(s)$ is a first order LPF. Similar to [14], $F_D(s)$ is defined as a fourth order HPF:

$$F_D(s)=\frac{g_bg_cg_dg_e}{(s+2\pi f_b)(s+2\pi f_c)(s+2\pi f_d)(s+2\pi f_e)} \tag{67}$$

We still let $\overline{\tau}=20$ s, $\Delta\tau$=0.1 s, and use the high gain PI controller ($g$=10000, $a$=100) as $G(s)$. Table 1 gives an example of filter parameters, with which criterion (62A) is satisfied, while $|H_M|_{max}=1+1.539\times10^{-5}$, is very close to 1.

Table 1. The filter parameters of a modified-dual arm locking sensor.

| Pole frequency | gain |
|---|---|
| $f_a$=0 Hz | $g_a$=3.3 |
| $f_b$=0.9 μHz | $g_b$=1 |
| $f_c$=0.9 μHz | $g_c$=1 |
| $f_d$=0.9 μHz | $g_d$=1 |
| $f_e$=0.9 μHz | $g_e$=1 |

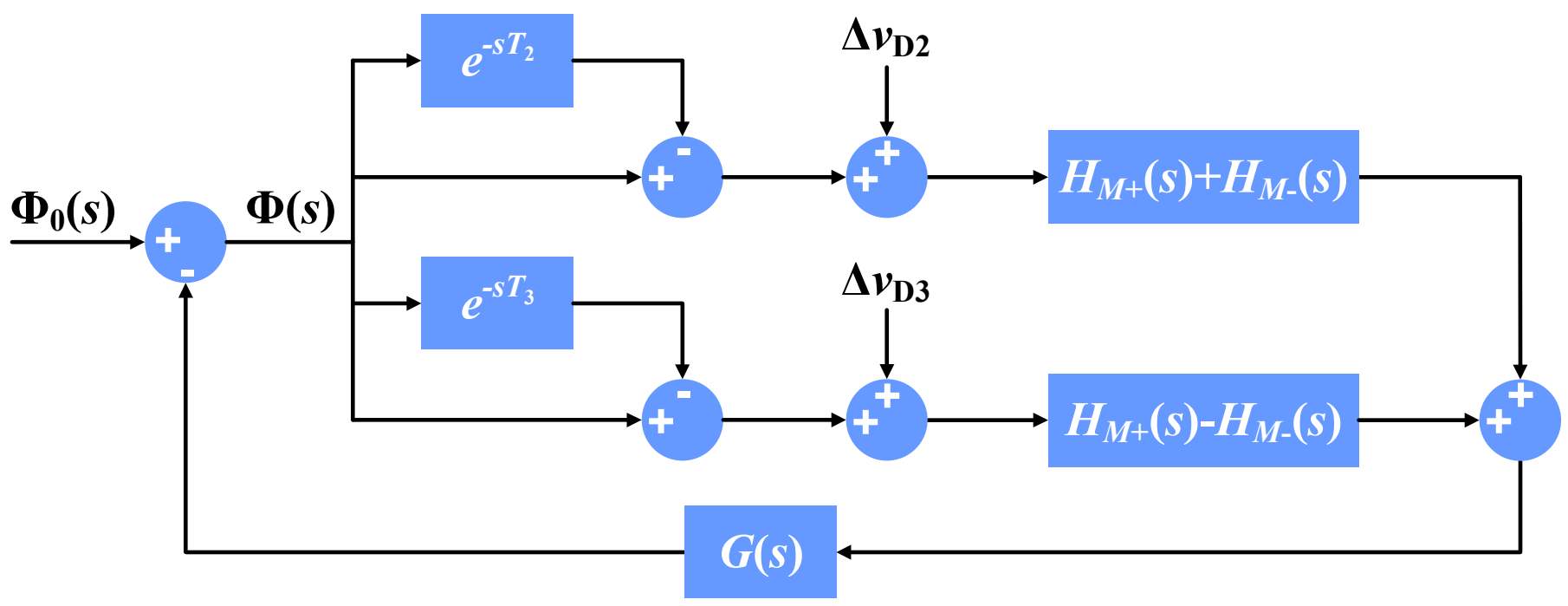

Figure 12. diagram of dual arm locking system with Doppler frequency errors.

Fig. 12 gives a general dual arm locking system with Doppler frequency errors $\Delta\nu_{D2}$ and $\Delta\nu_{D3}$. Based on this diagram, we have:

$$\Phi(s)=\frac{1}{H_D}\Phi_0(s)-H_{\nu+}\left(\Delta\nu_2+\Delta\nu_3\right)-H_{\nu-}\left(\Delta\nu_2-\Delta\nu_3\right) \tag{68}$$

where

$$H_D=1+\left(2-e^{-sT_2}-e^{-sT_3}\right)H_{M+}G+\left(e^{-sT_3}-e^{-sT_2}\right)H_{M-}G \tag{69}$$

$$H_{v+} = \frac{H_{M+}G}{H_D} \tag{70}$$

$$H_{v-} = \frac{H_{M-}G}{H_D} \tag{71}$$

Fig. 13 gives the amplitude response of $H_{v+}(s)$ and $H_{v-}(s)$. It can be seen that $H_{v+}(s)$ dominates at low frequencies below $10^{-6}$ Hz (Doppler shift oscillation frequency is at ~$10^{-7}$ Hz). And at low frequency limit, $H_{v+}(s)$ approximates to $1/(s2\overline{\tau})$. So, the Doppler frequency pulling rate is about $1/2\overline{\tau}$, which meets expectations of modified-dual arm locking.

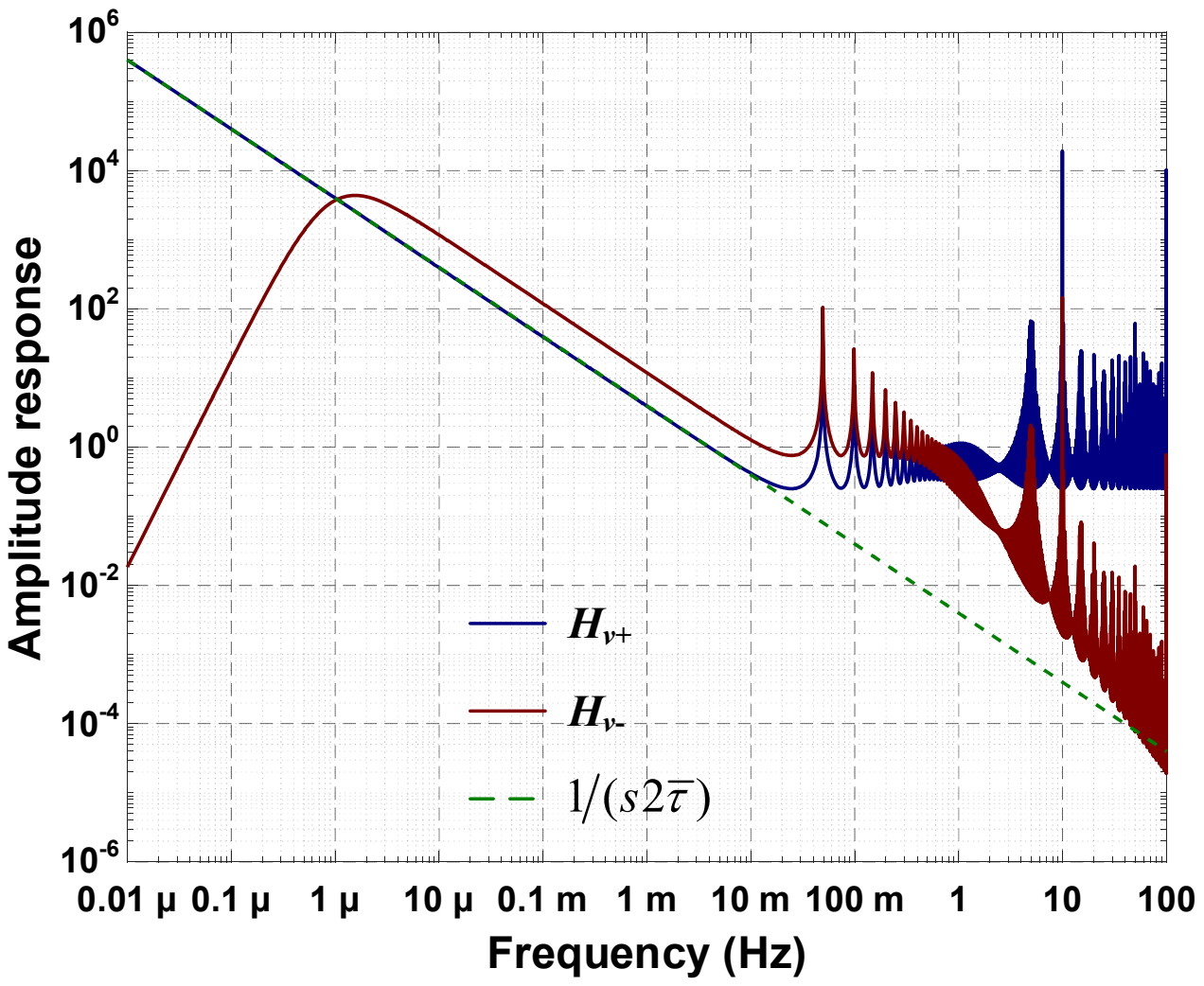

Figure 13. The amplitude response of $H_{v+}(s)$ and $H_{v-}(s)$.

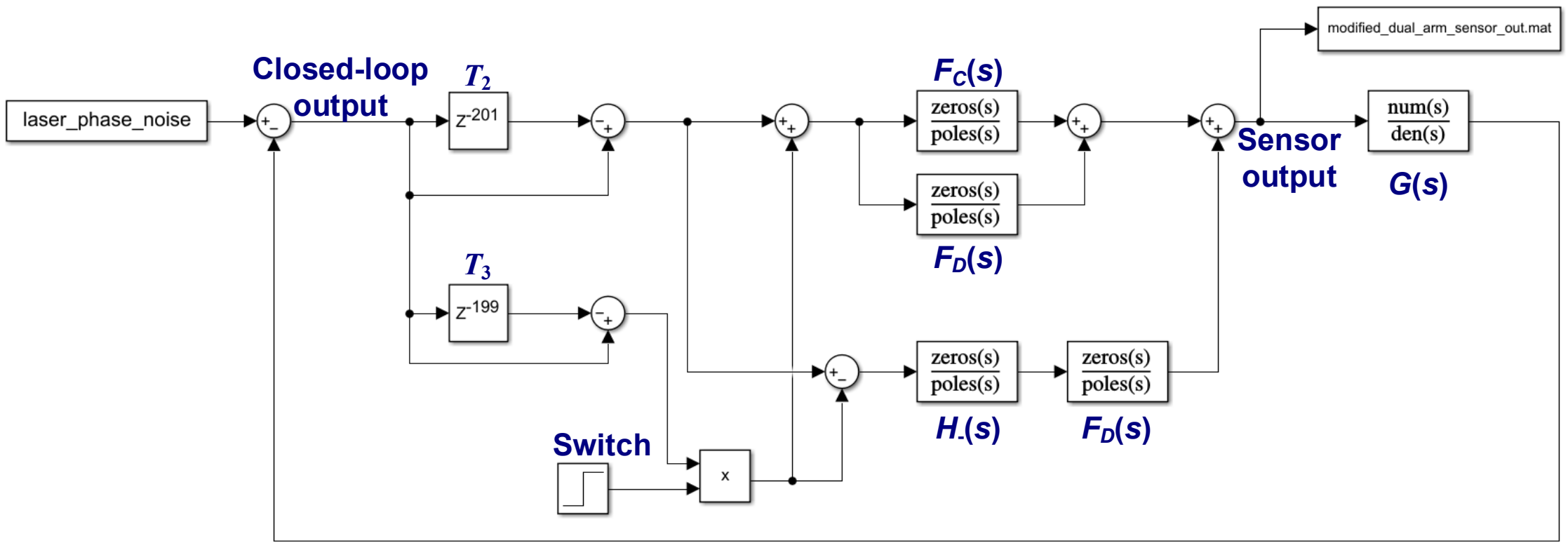

Figure 14. Simulink diagram of modified-dual arm locking with a PI controller, $G(s)$=10000($s$+100)/$s$.

Using the diagram of Fig. 14, modified-dual arm locking is also simulated in time domain over $10^5$ s by Simulink. The results are given in Fig. 15. The sensor output is still used as the evaluation metrics. Since the modified-dual arm sensor we use can amplify the laser phase noise by more than 100 times, to make the comparison more objective, we compare the sensor output of closed-loop with the original laser phase noise in Fig. 15. The RMS laser phase noise is reduced from $4.37\times10^7$ rad (original laser phase noise) to 0.0691 rad (closed loop sensor output). The transient response peaks decay from ~4.5 rad at the beginning (Fig. 15 (a) inset) to about ~0.04 rad at $10^5$ s (Fig. 15 (a) upper part). In Fig. 15(b). The laser phase noise at the sensor output is suppressed by ~8 orders of magnitude at 0.1 mHz, and ~4 orders around 1 Hz, relative to the original laser phase noise.

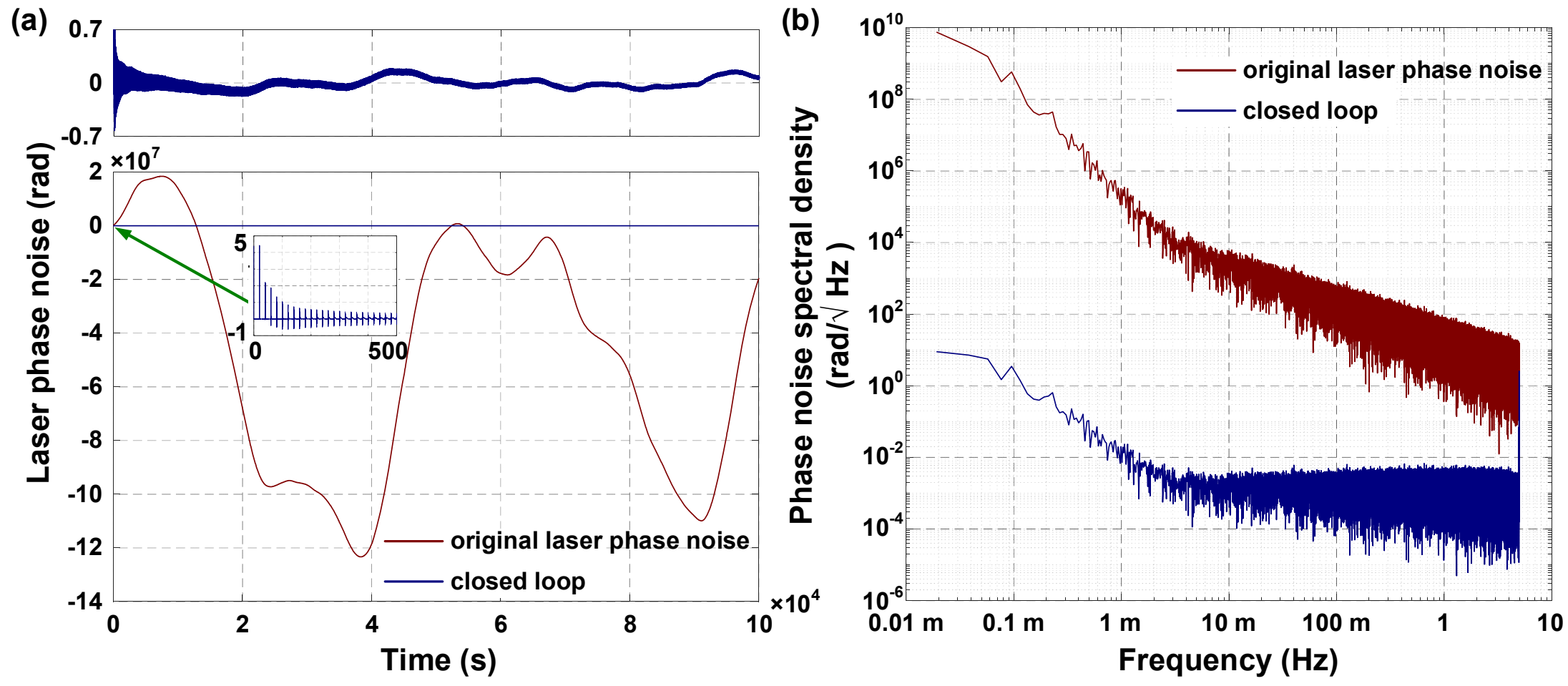

Figure 15. Modified-dual arm locking simulation results using a PI controller. (a) laser phase noise in time domain; (b) laser phase noise spectral density.

## 5. Common arm locking controller

The basic idea of modified-dual arm locking is to combine the advantages of low Doppler pulling rate of common arm locking and flat noise suppression in science band of dual arm locking. If we only need large noise suppression ratio in science band as well as low Doppler pulling rate, common arm locking itself may be enough.

Common arm locking can be obtained by setting $H_{+}(s) = 1$ and $H_{-}(s) = 0$ in Fig. 9. Therefore, Eq. (53) can be rewritten as

$$H_{CL}(s) = H_{0,C}\sum_{k=0}^{n} H_C^k e^{-sk\bar{\tau}}, \quad nT_2 \le t < mT_2 + lT_3 \tag{72}$$

where

$$H_{0,C}(s) = \frac{1}{1+2G(s)} \tag{73}$$

$$H_C(s) = \frac{G(s)}{1+2G(s)}\left(e^{-s\Delta\tau} + e^{s\Delta\tau}\right) \tag{74}$$

Then the stability criterion for a common arm locking controller can be obtained:

I. All the poles of $H_C(s)$ have negative real parts:

$$\mathrm{Re}(p_{i,C}) < 0, \quad i = 1, 2\cdots, r \tag{75A}$$

II. The amplitude response of $H_C(s)$ is always not higher than 1:

$$\max_{\omega\in[-\infty,+\infty]} \left|H_C(j\omega)\right| \le 1 \tag{75B}$$

It can be verified that the stability criterion (75A) and (75B) are automatically satisfied for a PI controller. the diagram of Fig. 16 is used to evaluate the performance of common arm locking with a high gain PI controller. The sensor output for open loop and closed loop are compared in Fig. 17. The RMS laser phase noise is reduced from $1.085\times10^6$ rad (open loop) to 0.362 rad (closed loop). The transient response peaks decay from ~36 rad at the beginning (Fig. 17 (a) inset) to about ~0.24 rad at $10^5$ s (Fig. 17 (a) upper part). Except for the residual peaks around 0.05 Hz and 1 Hz, the laser phase noise suppression radio is between $10^4$~$10^7$ within the science band.

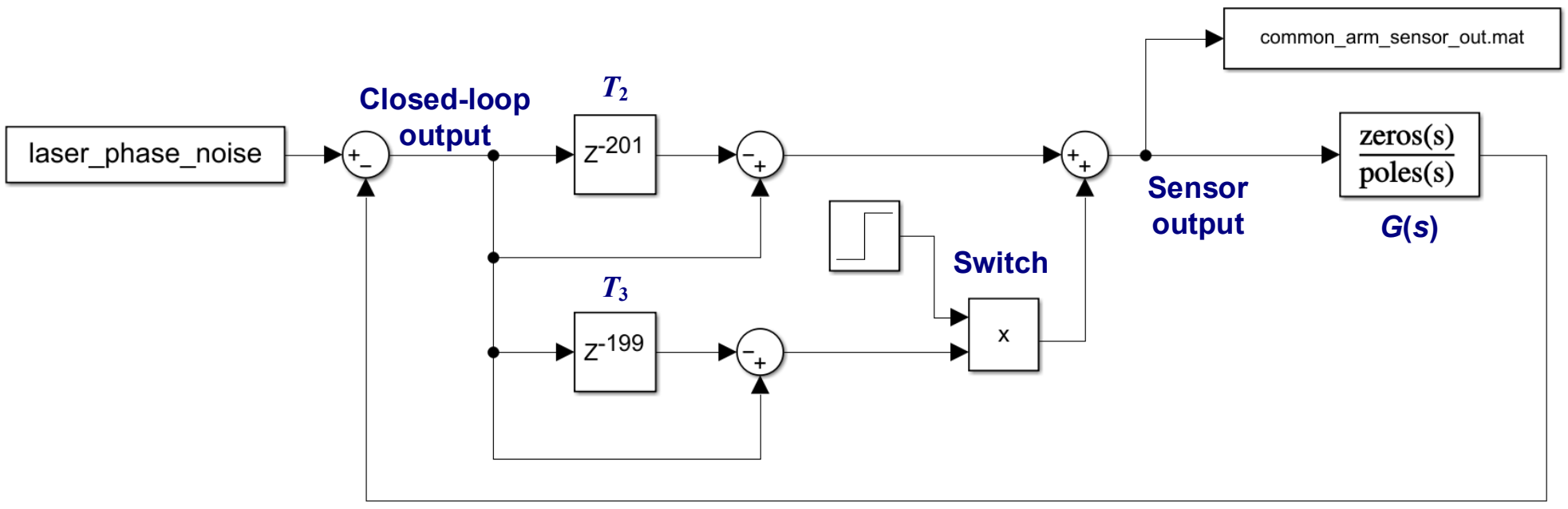

Figure 16. Simulink diagram of common arm locking with PI controller, $G(s)=10000(s+100)/s$.

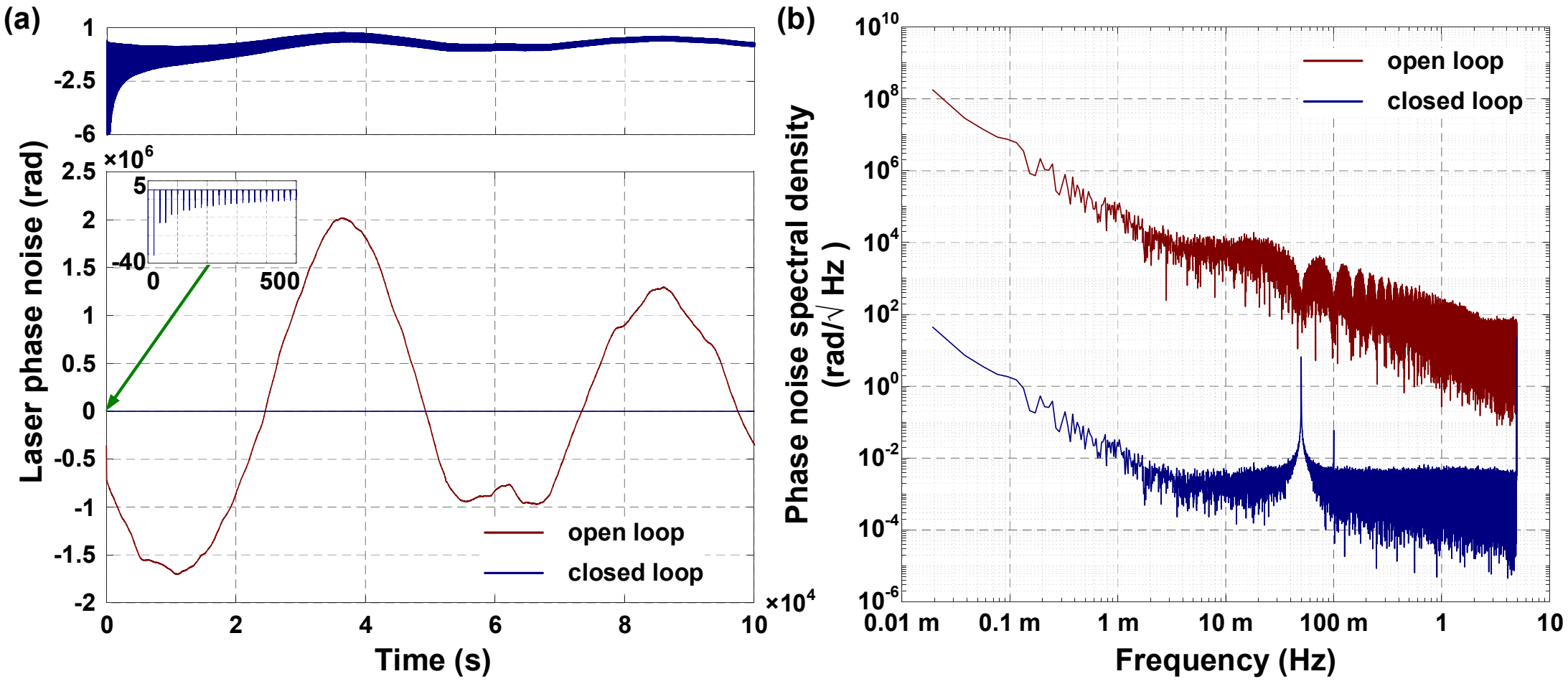

Figure 17. Common arm locking simulation results using PI controller. (a) laser phase noise in time domain; (b) laser phase noise spectral density.

To further compare the transient response of dual arm, modified-dual arm and common arm locking, the same input laser phase noise is used in the diagram of Fig. 10, 14 and 16, respectively, and the output data at the closed-loop output port of each setup is recorded. For short-term results in Fig. 18 (a), the three transient responses coincide in the first 25 s because $T_3$ arm is turned off. After 25 s, the dual arm responses quickly decay to zero due its inherent differential characteristics. While the other two cases decay very slowly due to the common arm contribution. For long-term results shown in Fig. 18 (b), the common arm has the lowest decay rate while the other two cases are almost equal. However, there is an additional long-term fluctuation imposed on the modified-dual arm locking's results, compared with dual arm case. This drift is actually caused by the larger than 1 part in $|H_M|$, which can accumulate to be a huge value after sufficiently long time and then break the locking.

HPFs were used in [14, 20, 21] to decouple the Doppler frequency noise at ~$10^{-7}$ Hz. However, our calculations showed that HPFs higher than order 2 may break the stability condition (75B). One possible solution to solve this problem is to increase the Doppler frequency estimation accuracy with emerging new technologies [34, 35]. If a real time Doppler frequency estimation at ~$10^{-7}$ Hz is performed, the residual Doppler frequency error after estimation is still at $10^{-7}$ Hz. With the help of low order HPFs, it is only necessary to guarantee that the in-loop Doppler frequency pulling will not exceed the PZT tuning range of the laser, and the Doppler frequency pulling can be filtered out by out-of-loop high order HPFs, which does not affect the system's stability.

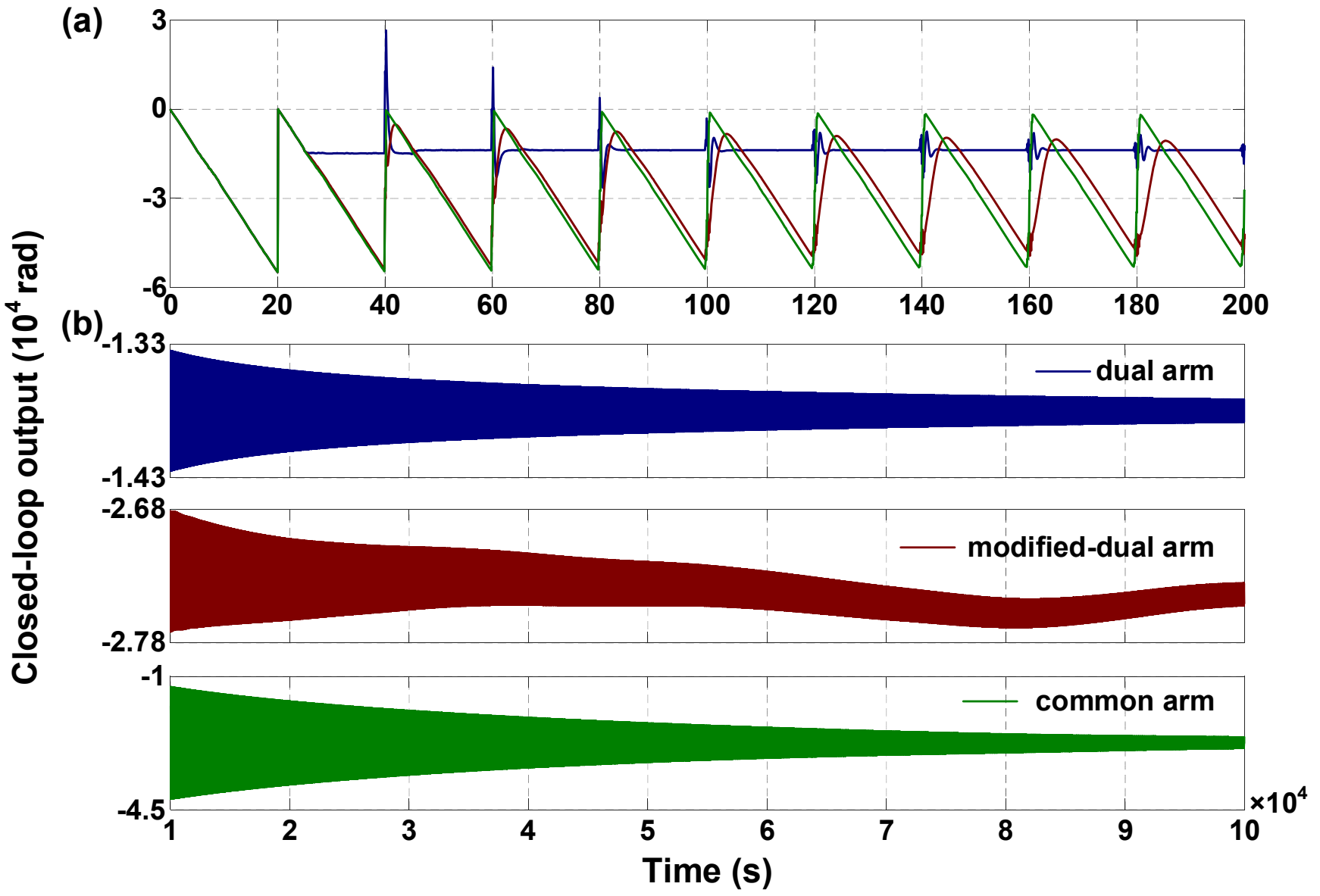

Figure 18. Transient response of dual arm, modified-dual arm and common arm locking systems. (a) short-term results; (b) long-term results. (a) shares the same legend in (b).

## 6. Other controllers

Since all the arm locking controllers reported so far were designed using the phase margin criterion, it is also interesting to test these controllers with the new criterions we just proposed.

The first one is a single arm controller given by Table 1 of [19]. Since it is designed for Taiji Project, let $T$=20 s. The characteristic equation of $H(s)$ defined by Eq. (23) can be written as

$$\sum_{k=0}^{7} c_k s^k = 0 \tag{76}$$

The coefficients $c_k$ ($k$=0, …, 7) and the roots of Eq. (76) are given in Table 2. Since all the roots have negative real parts, criterion (24A) is satisfied.

Table 2. The coefficients and roots of the characteristic equation of $H(s)$ from [19].

| Coefficients | Roots |
|---|---|
| $c_0$=1 | $p_1$=-280.3664 |
| $c_1$=111.11 | $p_2$=-29.4497 |
| $c_2$=3553.211 | $p_3$=-2.9798 |
| $c_3$=91232.421 | $p_4$=-0.2925 |
| $c_4$=304230.291 | $p_5$=-0.0136 + 0.0288i |
| $c_5$=102122.201 | $p_6$=-0.0136 - 0.0288i |
| $c_6$=3444.4 | $p_7$=-0.0124 |
| $c_7$=11 | |

The amplitude response of $H(s)$ is shown in Fig. 19. Since $|H|_{\max}$=1.0007>1, criterion (24B) is not satisfied. Thus, this controller is transient instable.

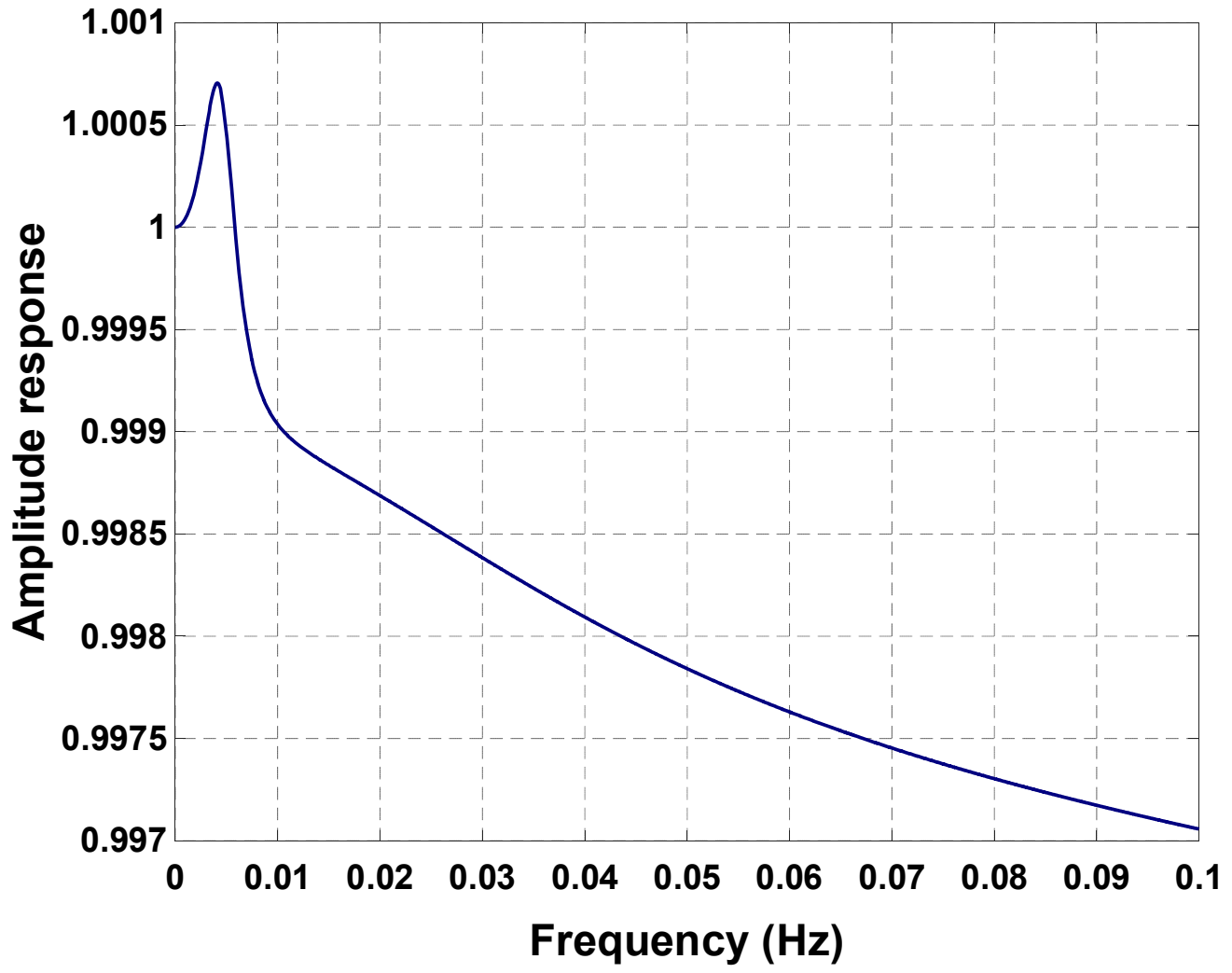

Figure 19. The amplitude response of $H(s)$ from [19].

To confirm our conclusion, the diagram of Fig. 20 is used to simulate the step response of this controller in a single arm locking system. The results are given in Fig. 21. Because $|H|_{\max}$ is very close to 1, in the first ~800 s, the transient responses decrease with time. Then the greater than 1 part of $H(s)$ will build up by the power of $H(s)^n$, so as to increase the transient response. In the whole simulation time of $2\times10^5$ s, the maximum value of the step response is larger than 2 (twice of the input level), thus this controller cannot be used to suppression laser phase noise.

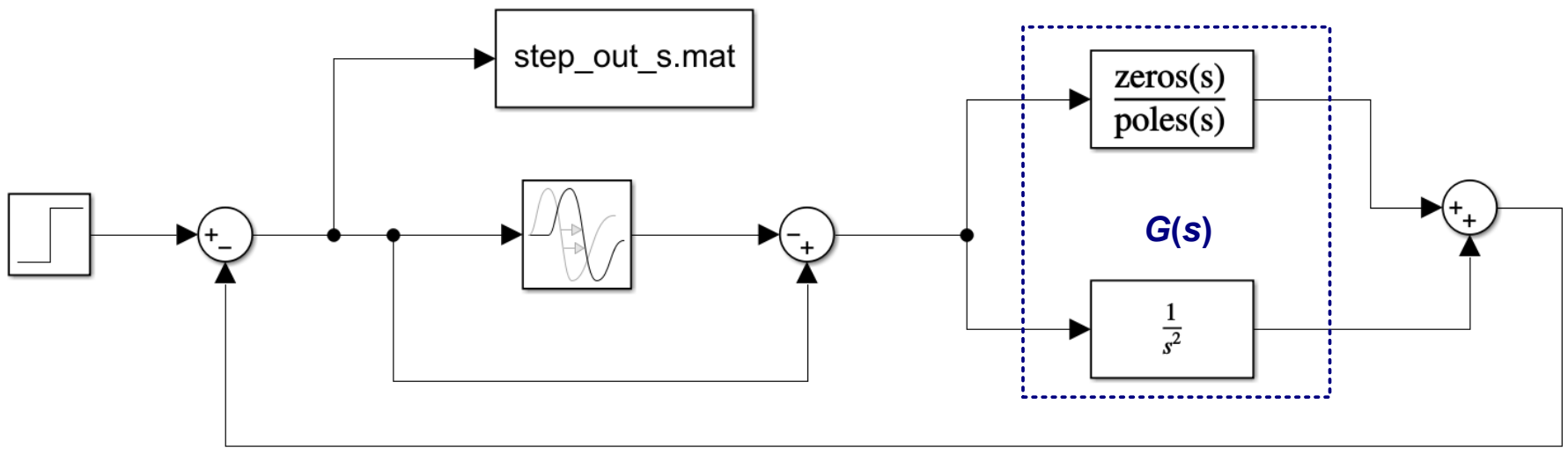

Figure 20. Simulink diagram of the single arm locking system from [19].

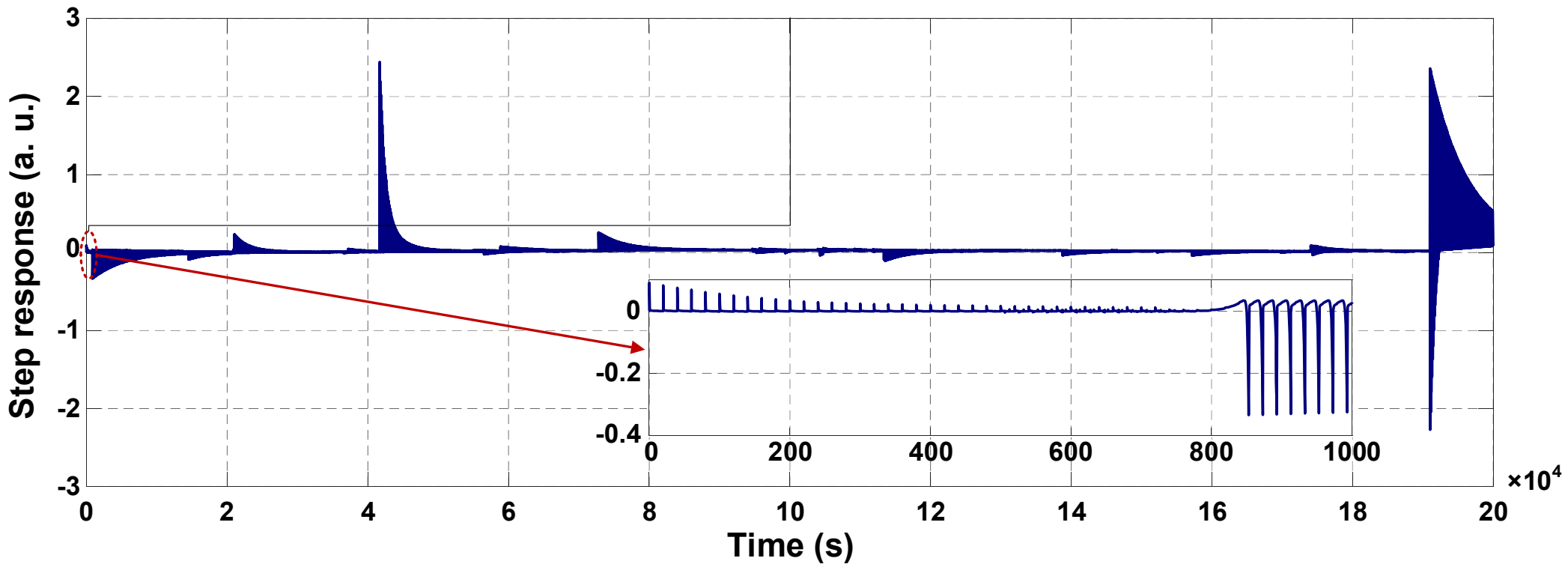

Figure 21. Step response of the single arm locking system from [19].

The second controller is a modified-dual arm controller from [20]. To be consistent with [20], let $\overline{\tau} = 16.6$ s, $\Delta\tau$=0.062 s. $H_{M+}$ and $H_{M-}$ in Eq. (61) are given by Eq. (21) and (22) of [20], and $G(s)$ is given by Eq. (31) of [20]. The characteristic equation of $H_M(s)$ can be written as

$$\sum_{k=0}^{28} c_k s^k = 0 \tag{77}$$

where $c_k$ ($k$=0, …, 28) are given by Table 3. The roots of this equation $p_i$ ($i$=1, …, 28) are also shown in Table 3. Since Re($p_{27}$)>0, Re($p_{28}$)>0, criterion (62A) is not satisfied.

Table 3. The coefficients and roots of the characteristic equation of $H_M(s)$ from [20].

| **Coefficients** | | **Roots** | |
|---|---|---|---|
| $c_0=0$ | $c_{15}=3.5057\times10^{33}$ | $p_1=0$ | $p_{15}=-0.0719$ |
| $c_1=1.2791\times10^{-15}$ | $c_{16}=9.8738\times10^{33}$ | $p_2=-4.0764\times10^{6}$ | $p_{16}=-0.0227 + 0.0668i$ |
| $c_2=3.8210\times10^{-9}$ | $c_{17}=8.6234\times10^{33}$ | $p_3=-5.2779\times10^{5}$ | $p_{17}=-0.0227 - 0.0668i$ |
| $c_3=3.8087\times10^{-3}$ | $c_{18}=2.5174\times10^{33}$ | $p_4=-8.5856\times10^{4}$ | $p_{18}=-0.0254$ |
| $c_4=6.9034\times10^{6}$ | $c_{19}=3.0751\times10^{32}$ | $p_5=-1.6511\times10^{4}$ | $p_{19}=-0.0043$ |
| $c_5=7.8777\times10^{10}$ | $c_{20}=1.9208\times10^{31}$ | $p_6=-1.6300\times10^{3}$ | $p_{20}=-0.0041$ |
| $c_6=3.6756\times10^{14}$ | $c_{21}=5.9892\times10^{29}$ | $p_7=-253.3542$ | $p_{21}=-5.2069\times10^{-4}$ |
| $c_7=8.9293\times10^{17}$ | $c_{22}=4.9575\times10^{27}$ | $p_8=-166.0358$ | $p_{22}= (-4.8141 + 0.5855i)\times10^{-4}$ |
| $c_8=1.1851\times10^{21}$ | $c_{23}=1.3102\times10^{25}$ | $p_9=-9.4554 +11.1435i$ | $p_{23}= (-4.8141 - 0.5855i)\times10^{-4}$ |
| $c_9=8.1813\times10^{23}$ | $c_{24}=7.0510\times10^{21}$ | $p_{10}=-9.4554 -11.1435i$ | $p_{24}= (-4.0807 + 0.3964i)\times10^{-4}$ |
| $c_{10}=2.5393\times10^{26}$ | $c_{25}=4.6450\times10^{17}$ | $p_{11}=-14.1270$ | $p_{25}= (-4.0807 - 0.3964i)\times10^{-4}$ |
| $c_{11}=3.3552\times10^{28}$ | $c_{26}=5.2681\times10^{12}$ | $p_{12}=-3.7896$ | $p_{26}=-5.3961\times10^{-8}$ |
| $c_{12}=1.5783\times10^{30}$ | $c_{27}=9.4173\times10^{6}$ | $p_{13}=-1.2588$ | $p_{27}= (2.6708 + 5.2158i)\times10^{-8}$ |
| $c_{13}=3.2695\times10^{31}$ | $c_{28}=2$ | $p_{14}=-0.3461$ | $p_{28}= (2.6708 - 5.2158i)\times10^{-8}$ |
| $c_{14}=4.3478\times10^{32}$ | | | |

The amplitude response of $H_M(s)$ is shown in Fig. 22. It can be seen that $|H_M|_{\max}$=1.8173>1, so criterion (62B) is also not satisfied. Therefore, the modified-dual arm locking sensors and controller of [20] are instable.

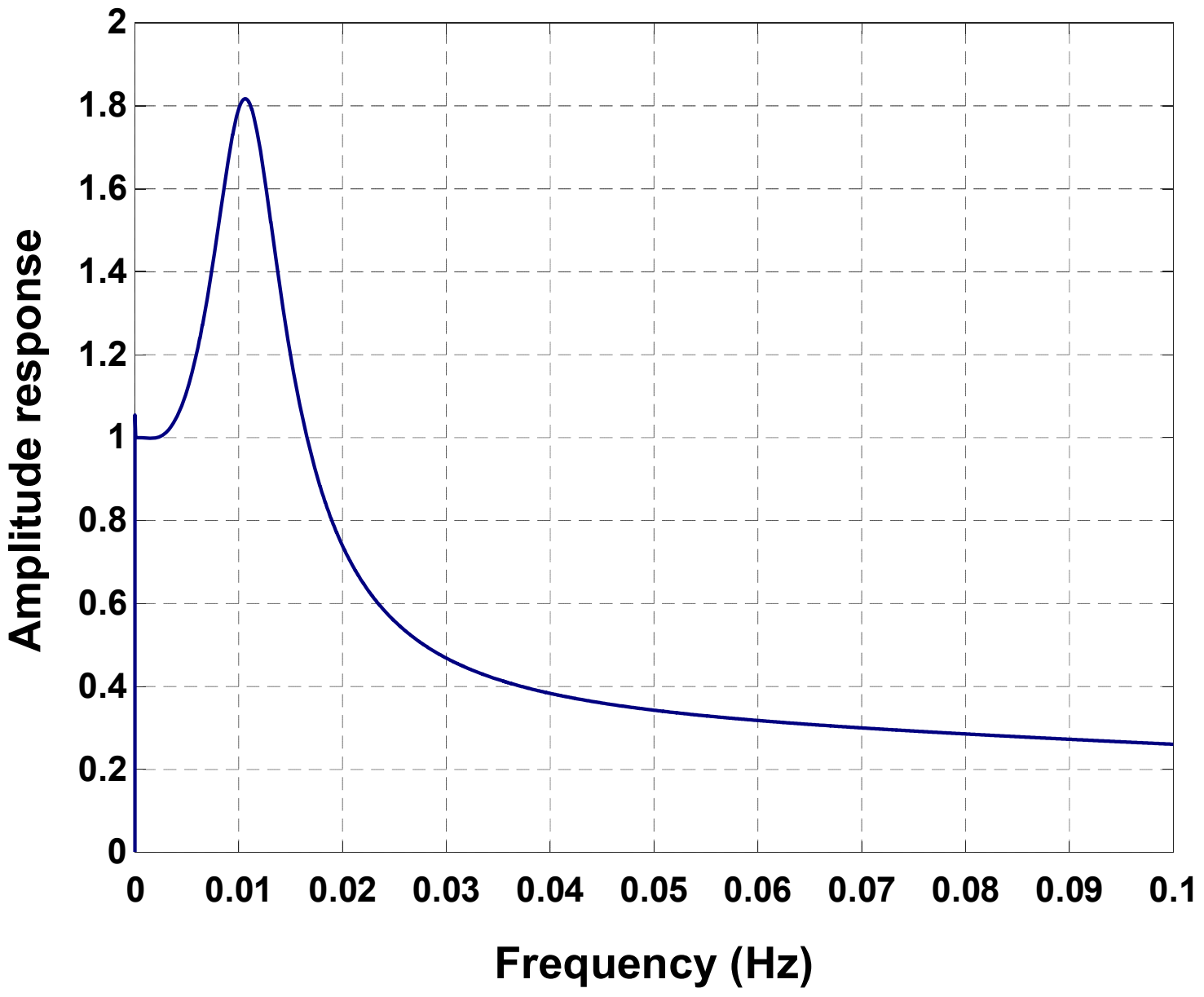

Figure 22. The amplitude response of $H_M(s)$ from [20].

To verify our statement, the diagram of Fig. 23 is used to simulate the step response of the modified-dual arm locking system of [20]. The results are shown in Fig. 24. It can be seen that the response becomes divergent after 2000 s, and after $10^5$ s, the response is larger than 1 (input level).

Both of the two examples above indicate that phase margin criterion cannot guarantee the stability of an arm locking system. Actually, phase margin criterion is an empirical method in engineering, there is no evidence that it works for all feedback systems. The criterions we propose in this paper is based on rigorous mathematical derivation, they are applicable for any feedback systems with parallel delay modules.

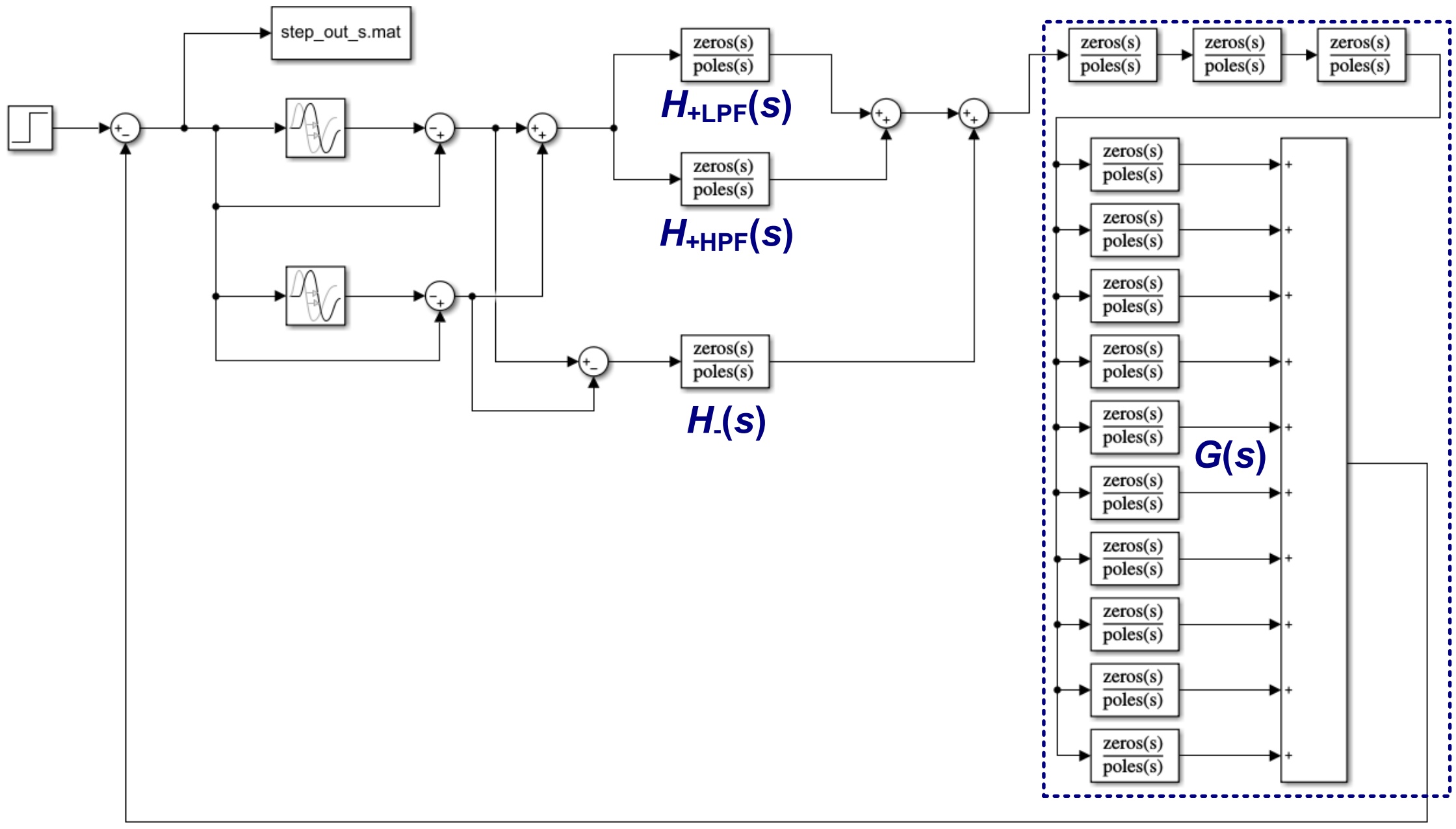

Figure 23. Simulink diagram of the modified-dual arm locking system from [20].

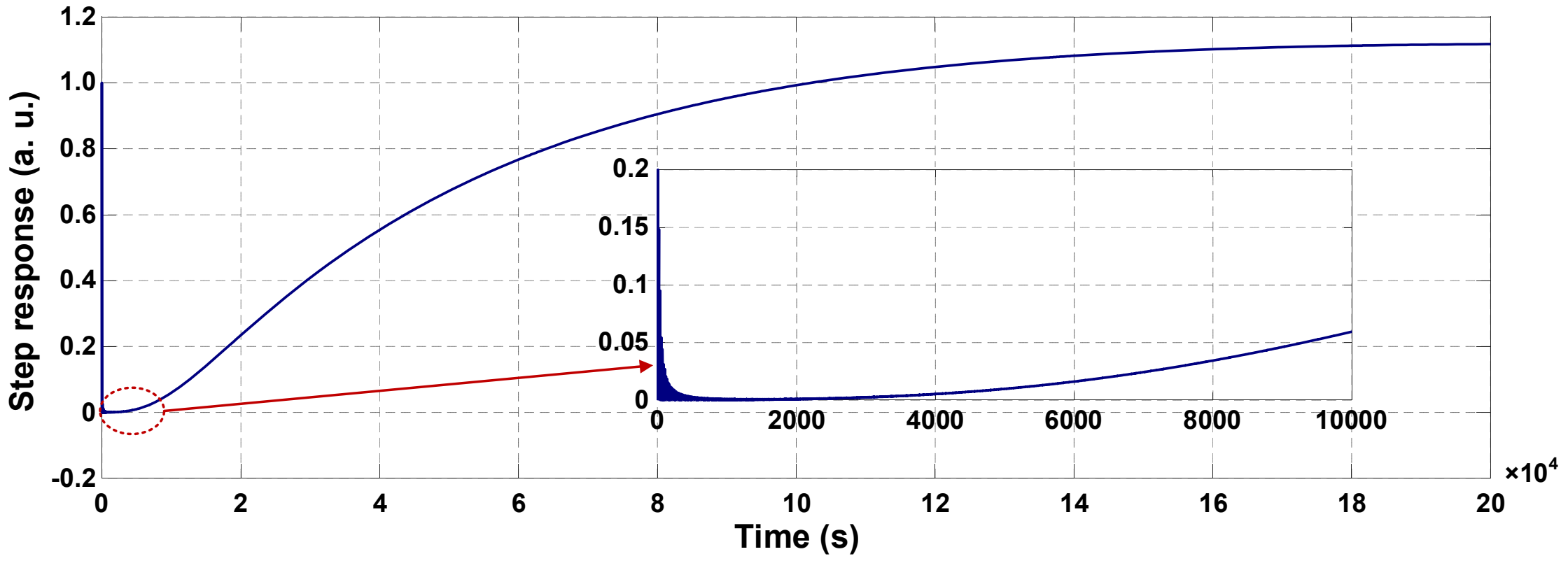

Figure 24. Step response of the modified-dual arm locking system from [20].

## 7. Conclusion

In conclusion, a comprehensive transient analysis is carried out for arm locking systems. Analytical stability criterions for single arm locking controllers are given by rigorous mathematical derivation. These criterions can be extended to general arm locking architectures, such as dual arm locking, modified-dual arm locking and common arm locking. Using these criterions, the design rules for different kinds of arm locking systems are provided. In most cases, PI controller can automatically meet these stability criterions, and the preliminary simulation results in time domain show that a high gain PI controller may be enough to suppress the laser phase noise by 5 orders of magnitude within the science band. To keep the system stable, it would be better to filter out other noise sources, such as Doppler pulling, at out of the loop rather than in the loop, so as to let the controller focus on the laser noise suppression task.

With our stability criterions and Simulink verifications, we also find that the widely-used phase margin criterion cannot guarantee the arm locking stability, which means most of arm locking controllers reported so far may have potential instability. Therefore, our work is significant in terms of the arm locking design strategies. Besides, the stability criterions in this paper can also be used in other feedback systems, where several modules with different delays are connected in parallel.

## 8. Acknowledgement

This work is supported by National Key R&D Program of China (Grant No. 2021YFC2201902 and 2021YFC2201901) and National Natural Science Foundation of China (Grant No. 61975149). The authors thank Guoqing Chang for valuable discussions.

## Appendix

**Theorem 1.** (A.1) is an equation of degree $n$ in variable $x$:

$$x^n + a_{n-1}x^{n-1} + \cdots + a_1 x + a_0 = 0 \tag{A.1}$$

where $a_k$ ($k$=0, 1, …, $n$-1) are real coefficients. If all the roots of (A.1) have negative real parts, then $a_k$>0 ($k$=0, 1, …, $n$-1).

Proof: Suppose (A.1) have $m$ real roots $r_k$ ($k$=1, …, $m$) and $l$ (=($n$-$m$)/2) pairs of complex roots $p_k \pm iq_k$ ($k$=1, …, $l$), then:

$$x^n + a_{n-1}x^{n-1} + \cdots + a_1 x + a_0 = \prod_{k=1}^{l}(x^2 + b_k x + c_k)\prod_{k=1}^{m}(x + d_k) \tag{A.2}$$

If $r_k<0$ ($k=1, \ldots, m$), $p_k<0$ ($k=1, \ldots, l$), then $b_k=-2p_k>0$, $c_k=p_k^2+q_k^2>0$ ($k=1, \ldots, l$), $d_k=-r_k>0$, ($k=1, \ldots, m$). Expanding the right side of (A.2), we have $a_k>0$ ($k=0, 1, \ldots, n-1$).